\documentclass[graybox, secnum]{svmult}

\usepackage{mathptmx}       
\usepackage{helvet}         
\usepackage{courier}        
\usepackage{type1cm}        
\usepackage{makeidx}         
\usepackage{graphicx}        
\usepackage{multicol}        
\usepackage[bottom]{footmisc}
\usepackage{hyperref}        
\usepackage{soul}            
\usepackage{xcolor}
\definecolor{bluecite}{HTML}{0875b7}

\hypersetup{
	colorlinks=true,
	linkcolor=bluecite,
	urlcolor=bluecite,
	citecolor=bluecite
}
\usepackage[square,numbers,compress]{natbib}
\makeindex             

\usepackage{amsmath,amsfonts,amssymb}
\usepackage{subcaption}
\usepackage{bbold}
\usepackage{tensor,slashed}
\usepackage[T1]{fontenc}
\usepackage[utf8]{inputenc}

\newcommand{\be}{\begin{equation}}
\newcommand{\ee}{\end{equation}}

\renewcommand{\imath}{\ensuremath{\mathrm{i}}}

\newcommand{\p}{\ensuremath{\partial}}

\newcommand{\cM}{\mathcal{M}}

\usepackage{ulem}

\begin{document}
\title*{Stability properties of Regular Black Holes}
\author{Alfio Bonanno and Frank Saueressig}
\institute{Alfio Bonanno  \at INAF, Osservatorio Astrofisico di Catania,
	via S.Sofia 78, I-95123 Catania, Italy; \newline
	INFN, Sezione di Catania, via S.Sofia 64, I-95123, Catania, Italy \newline \email{alfio.bonanno@inaf.it}
	\and Frank Saueressig  \at  Institute for Mathematics, Astrophysics and Particle Physics (IMAPP), Radboud University, \newline Heyendaalseweg 135, 6525 AJ Nijmegen, The Netherlands \newline \email{f.saueressig@science.ru.nl}}
\maketitle

\abstract{Black holes encountered in general relativity are characterized by spacetime singularities hidden within an event horizon. These singularities provide a key motivation to go beyond general relativity and look for regular black holes where the spacetime curvature remains bounded everywhere. A prominent mechanism achieving this replaces the singularity by a regular patch of de Sitter space. The resulting regular geometries exhibit two horizons: the outer event horizon is supplemented by an inner Cauchy horizon. The latter could render the geometry unstable against perturbations through the so-called mass-inflation effect, i.e., an exponential growth of the mass function. This chapter reviews  the mass-inflation effect for spherically symmetric black hole spacetimes contrasting the dynamics of the mass function for Reissner-Nordst{\"o}m and regular black holes. We also cover recent developments related to the late-time attractors induced by Hawking radiation which exorcise the exponential growth of the spacetime curvature encountered in the standard mass-inflation scenario. In order to make the exposition self-contained, we also briefly discuss basic properties of regular black holes including their thermodynamics.}

\newpage
\section{Introduction}
\label{sect.1}
To an external observer black holes are extremely simple objects. In the wake of a gravitational 
collapse quadrupole moments and all the deformations produced by the star either get swallowed inside the event horizon or are carried away by gravitational radiation. At late times, the external field settles into a Kerr-Newman geometry and is completely described by its mass, charge, and angular momentum.

The stationary exterior field hides a rather complicated dynamics in the black hole interior which drives spacetime towards its final classical fate, determined by Penrose's celebrated 1965 theorem \cite{Penrose:1964wq}. The key property of the black hole interior is that the Schwarzschild radial coordinate $r$ becomes time-like within the event horizon. Thus {\it a descent into  a black hole is a progression in time}. The inner layers do not only enclose but actually {\it precede} the core. This peculiar causal structure allows us to theoretically explore the interior by boring in ``layer-by-layer'' starting from shells situated at larger radii. The causal structure guarantees that  what we learn about the outer zone, where the classical description of the geometry is still possible, cannot be affected by our ignorance about quantum gravity effects potentially operating in the innermost regions close to the spacetime singularity. From this point of view, it is possible to speak of an ``evolution'' with increasing advanced time and subsequent relaxation to its final state. The final ``big-crunch'' can then be imagined as an abrupt stop or a rupture in the classical geometry as, accordingly to the strong cosmic censorship, the inner singularity is space-like \cite{Penrose:1969pc}.

In the case of no angular momentum  the static configuration approached outside the horizon is mirrored in the interior by an almost spherically symmetric configuration where the internal disturbances are exponentially damped. As the radial coordinate plays the role of time, it is conceivable that, before the singularity, there exists a layer where the dynamics admits a semiclassical description in terms of an effective 
Einstein equation originating from quantum field theory in a curved spacetime \cite{Birrell:1982ix}
\begin{equation}
	\label{efe}
	G_{\mu\nu}= 8 \pi G \, \langle T_{\mu\nu} \rangle \, . 
\end{equation}
Here $G_{\mu\nu}$ is the (classical) Einstein tensor, $G$ is Newton's coupling, and $\langle T_{\mu\nu} \rangle$ is the (suitably regularized) expectation value of the stress-energy tensor. Using Schwarzschild coordinates, the sign of $\langle {T_t}^t \rangle$ dictates  the fate of the singularity. If it is negative, quantum polarization effects have a self-regulatory effect. The Misner-Sharp quasi-local mass, formally introduced in eq.\ \eqref{def.MisnerSharp}, behaves like $ M(r) \sim r^3$ near Planckian distances. This in turn implies a de Sitter core inside the  black hole \cite{Poisson:1988wc}. 

The idea of replacing the spacetime singularity by a patch of de Sitter space actually comes with a long history. 
It was proposed in \cite{bardeen1968non}, introduced in \cite{1982ZhPmR..36..214M} in the context of the limiting-curvature conjecture and further developed in \cite{dy92,frolov98}, also see \cite{ansoldi} for a review. The formation and evaporation of these regular black holes  has been first discussed in \cite{hayward}.
From a fundamental perspective these types of regular black holes have been motivated based on the gravitational asymptotic safety program \cite{Bonanno:2000ep,Bonanno:2006eu} (also see \cite{Cai:2010zh,Reuter:2010xb,Koch:2013owa,Koch:2014cqa,Torres:2014gta,Bonanno:2017zen,Pawlowski:2018swz,Adeifeoba:2018ydh,Platania:2019kyx} for selected follow-up works) and as Planck stars \cite{Rovelli:2014cta} (further explored in \cite{Barrau:2014hda,DeLorenzo:2014pta,Christodoulou:2016vny} and reviewed in \cite{Perez:2017cmj}) inspired by loop quantum gravity.

The de Sitter core shared by these models 
has important consequences for their causal structure, 
since it implies the presence of an inner horizon, a so-called Cauchy horizon. This feature may be crucial when considering the regularity of the geometry in the presence of perturbations. Already in 1968 Penrose noted \cite{penrose68} that the ingoing sheet of the inner horizon, corresponding to infinite advanced time, is a surface of infinite blueshift for a wavelike disturbance propagating inwards. In particular, the radiative tail of a generic collapse  experiences an exponential blueshift close to the Cauchy horizon. Most likely, this property signals the presence of an instability.  

This possibility was then explored in several perturbative studies. The first investigation of the backreaction of the blueshifted influx onto the geometry near the Cauchy horizon was 
performed by Hiscock \cite{1981PhLA...83..110H}, but it was 
Poisson and Israel \cite{poisson89} and Ori \cite{ori91} 
who demonstrated that the combined effect of influx and outflux produces an exponential growth of the Coulomb component
of the Weyl curvature as a function of the advanced time coordinate $v$. This has been dubbed the ``mass-inflation effect''. For an astrophysical (non-zero angular momentum) black hole this effect has dramatic consequences: at variance with the strong  singularity at the center, the curvature reaches planckian levels even if the radius of the inner horizon is macroscopically large. Semiclassical quantum corrections on a dynamically inflating geometry cannot really halt the build-up of this singularity. It turns out though that the new singularity is weak in the Tipler sense. For this reason, some authors have speculated that a $C^1$-extension of the geometry is not ruled out \cite{Burko:1995uq}. 
The arguments suggesting a possible continuation of spacetime beyond the Cauchy horizon carry over to the case of a regular black hole. In this case, the Cauchy horizon may be located at planckian distances from the center and only a complete and consistent theory of quantum gravity may be able to deliver a final answer on the possibility of determining plausible extensions of the geometry though.

The relevant question is if it is possible to identify  a  {\it dynamical   mechanism} damping of the mass-inflation singularity which operates near the central nucleus, despite our ignorance of the details concerning the structure of the spacetime near the center. 
In fact, the late advanced time geometry can be described in clearer physical terms because the initial data is well known in this case. It is composed of two elements:  the influx of gravitational waves transmitted by the outer potential barrier inside the black hole and the Hawking flux. The decay of the mass associated with the former is well-described by an inverse power law of the type $1/v^{p-1}$ \cite{price72a}, the so-called Price's tail ($p=12$ for a quadrupole moment). 
The latter contribution has been
conjectured in \cite{1991PhLA..161..223B} and it was argued that this should stop the growths of the mass function
at {\it early} advanced times already for mini black holes of mass $<100$ kg, approximately.

In the case of regular black holes the Hawking flux has a dramatic impact on  the final state of the geometry: it is given by an extremal configuration which is reached in an infinite amount of advanced time. The problem of the stability of the inner structure of a regular black hole then turns into a delicate dynamical question with two competing time scales, the one associated with the mass-inflation instability and the characteristic time-scale of the evaporation. If the latter is  much longer than the dynamical time-scale of the instability, one can completely neglect the effect of the Hawking radiation unless the black hole  is of very low mass. On the other hand
recent calculations showed \cite{Bonanno:2020fgp} that for specific types of regular black holes the exponential instability  is turned into a much milder power-law growth of the curvature near the Cauchy horizon. In this case one has to asses the impact of the
Hawking flux on the instability, using a dynamically evolving background which is running towards  its critical state in the $v\rightarrow\infty$ limit. This analysis was recently performed in \cite{Bonanno:2022jjp} and discovered two new classes of late-time attractors governing the dynamics of mass-inflation at asymptotically late times.

The goal of this chapter is to review the mass-inflation effect for regular black hole geometries in case of a static background as well as in the presence of the Hawking flux. In order to make the discussion self-contained we review the key properties of spherically symmetric spacetimes and regular black holes in Sects.\ \ref{sect.2.1} and \ref{sect.3} respectively. Sect.\ \ref{sect.2} covers the ``classical'' analysis of the mass-inflation effect for static Reissner-Nordstr{\"o}m black holes. Sect.\ \ref{sect.4} contrasts this situation to the mass-inflation effect encountered for regular black holes. Our conclusion that \emph{regular black holes come with stable cores} is detailed in Sect.\ \ref{sect.5}. Throughout this chapter we work in the probe-approximation, neglecting the back-reaction of the geometry onto perturbations.

\section{Spherically symmetric spacetimes: a primer}
\label{sect.2.1}
We start by summarizing the relevant properties of spherically symmetric spacetimes. Throughout the discussion we adopt geometric units $G=c=1$. A line-element exhibiting spherical symmetry can be cast into the form
\be\label{eq:lineelement1}
ds^2 = g_{ab} dx^a dx^b + r^2 d\Omega^2 \, , \qquad a,b = 0,1 \, .
\ee
Here, $x^a$ is any pair of coordinates that labels the set of two-spheres,  $d\Omega^2 \equiv d\theta^2+\sin^2(\theta ) d\varphi^2$ is the metric on a unit sphere and $r(x^a)$ is a function of $x^a$ defined by the geometric condition that the area $A$ of the two-spheres is given by $A=4\pi r^2$. We then define $f(x^a)$ through the gradient
\begin{equation}\label{gmass}
	f(x^a) \equiv g^{ab}(\partial_a r)(\partial_b r) \, . 
\end{equation}
The function $f(x^a)$ is related to the Misner-Sharp mass $M(x^a)$ through the definition
\be\label{def.MisnerSharp}
f(x^a) = 1 - \frac{2M(x^a)}{r} \, . 
\ee
A convenient choice for the coordinates $x^a$ is ingoing Eddington–Finkelstein coordinates $x^a = (v,r)$ where \eqref{eq:lineelement1} takes the form
\be\label{eq:eddfink}
ds^2 =  - f(v,r) \, dv^2 + 2 dr dv + r^2 d\Omega^2 \, . 
\ee
We then use a dot to denote a derivative with respect to $v$, $\dot{f} \equiv \frac{\p f(v,r)}{\p v}$, while derivatives with respect to $r$ are indicated by primes, $f^\prime \equiv \frac{\p f(v,r)}{\p r}$. 

Starting from \eqref{eq:eddfink}, it is instructive to express the most common curvature scalars in terms of $M(v,r)$ and its derivatives. For the Ricci scalar $R$ one readily finds
\be\label{eq.Ricci}
R =   \frac{2 r M^{\prime\prime}+4 M^\prime}{r^2} \, .
\ee
In addition, we define the two curvature scalars appearing at second order of the spacetime curvature. The Kretschmann scalar $K \equiv R_{\alpha\beta\gamma\delta} R^{\alpha\beta\gamma\delta}$ and the square of the Weyl-tensor $C^2 \equiv C_{\alpha\beta\gamma\delta} C^{\alpha\beta\gamma\delta}$ evaluate to
\begin{equation}\label{eq.curvature}
	\begin{split}
		K= & \, \frac{16 M M^{\prime\prime}}{r^4}+\frac{4 (M^{\prime\prime})^2}{r^2}-\frac{64 M M^\prime}{r^5}+\frac{32 (M^\prime)^2}{r^4}-\frac{16 M^\prime M^{\prime\prime} }{r^3}+\frac{48 M}{r^6} \, , \\
		C^2 = & \frac{4 \left(r^2 M^{\prime\prime}-4r M^{\prime}+6 M\right)^2}{3 r^6} \, . 
	\end{split}
\end{equation}
Since the geometries considered in this work are in general not Ricci flat, $R_{\mu\nu} \not = 0$, $K$ and $C^2$ are not identical and capture different curvature properties. For later reference we also define the Coulomb-component $\Psi_2 \equiv - \frac{1}{2}{C^{\theta\varphi}}_{\theta\varphi}$ of the Weyl-tensor which evaluates to
\begin{equation}\label{Weylcurve}
	\Psi_2 = -\frac{M}{r^3}+ \frac{2 M^{\prime}}{3 r^2}-\frac{M^{\prime\prime}}{6 r} \, . 
\end{equation}

For static geometries, where $f(v,r)$ is a function of $r$ only, 
the location of the horizons is determined by the condition 
\be\label{horizon}
f(r)=0 \, ,  \qquad r > 0 \, .
\ee
For the generic situation discussed in this review, this equation has two solutions. The event horizon (EH) is located at $r_+$. In addition, there is a Cauchy horizon (CH) at $r_- < r_+$. The surface gravity $\kappa_\pm$ at these points is defined as
\be\label{surface-grav}
\kappa_\pm \equiv \pm \left. \frac{1}{2} \frac{\p f(r)}{\p r} \right|_{r = r_\pm} \, . 
\ee
The choice of sign ensures that $\kappa_\pm > 0$. In non-static spacetimes the solutions of \eqref{horizon} may depend on $v$ and thus constitute  apparent horizons.

We exemplify the general setting for the static Reissner-Nordstr{\"o}m (RN)-geometry. In this case, eq.\ \eqref{gmass} is given by
\be\label{RNgeo}
f(r) = 1 - \frac{2m}{r} + \frac{Q^2}{r^2} \, , 
\ee
where $m$ is the asymptotic mass of the configuration and $Q$ denotes the charge of the black hole. In this case the EH and CH are situated at
\be\label{RN.horizons}
r_\pm = m \pm (m^2 - Q^2)^{1/2} \, . 
\ee
The corresponding surface gravity is given by
\be\label{RN.surfacegrav}
\kappa_\pm = \frac{m^2-Q^2}{r_\pm^2} \, . 
\ee
For $Q=m$ the position of the horizons coincide and one has the extremal RN black hole. For this configuration, the surface gravity \eqref{RN.surfacegrav} vanishes.

\section{Regular black holes}
\label{sect.3}
Upon completing our review of spherically symmetric geometries, we briefly summarize the key properties of regular black holes. The idea of a regular de Sitter core is introduced in Sect.\ \ref{sect.31} and various models implementing this idea are reviewed in Sect.\ \ref{sect.32}. We close with a short discourse on the thermodynamical properties of regular black holes in Sect.\ \ref{sect.33}.
\subsection{de Sitter cores}
\label{sect.31}
The idea of a de Sitter core regularizing the interior of a Schwarzschild black hole has been implemented in various settings. It is therefore instructive to start by reviewing the general idea, also see \cite{Frolov:2016pav} for a general discussion. Let us consider a static, spherically symmetric geometry characterized by the Misner-Sharp mass $M(r)$ introduced in eq.\ \eqref{def.MisnerSharp}. The general expressions for the Ricci scalar $R$ and the Kretschmann scalar $K$ for this case have been given in eqs.\ \eqref{eq.Ricci} and \eqref{eq.curvature}. Demanding that the curvature scalars remain finite as $r \rightarrow 0$ implies that $M(r)$ cannot be constant. Regularity demands that
\be\label{eq.regularityM}
\lim_{r \rightarrow 0} \, r^{-3} M(r) = \text{const} \, . 
\ee
Assuming a polynomial expansion around $r=0$, regularity then implies
\begin{equation}
	M(r) \simeq a_3 \, r^3+ a_4 \, r^4 +a_5 \, r^5 + {\cal O}(r^5) \, , 
\end{equation} 
where $a_i$, $i=3,4,\cdots$ are real coefficients. Substituting this expansion into eq.\ \eqref{def.MisnerSharp} gives
\begin{equation}\label{eq.frasym}
	f(r) \simeq 1-2 a_3 \, r^2-2 a_4 \, r^3-2 a_5 \, r^4 + {\cal O}(r^5) \, . 
\end{equation}
For $a_3 > 0$ the local geometry at $r=0$ is the one of de Sitter space.\footnote{For $a_3 < 0$ one encounters a regular anti-de Sitter core. Regular geometries of this type have recently been discussed in the context of Gauss black holes \cite{Boos:2021kqe}. The loop black hole suggested in \cite{Modesto:2008im} also falls into this classification having $a_3 = 0$.} Combining the asymptotics \eqref{eq.frasym} with the condition of asymptotic flatness shows that a black hole with a de Sitter core must come with an even number of horizons. This result is independent on the adopted field equations. It follows from the analytical properties of $M(r)$ near the origin.  In static spacetimes with only two horizons, the second, inner horizon is a Cauchy horizon. For non-static spacetimes, the regularity condition implies that the apparent horizon cannot cross $r=0$.

\subsection{Examples of regular black hole geometries}
\label{sect.32}
Regular black holes with a de Sitter core have been constructed by several authors. Prominent examples are the Bardeen black hole \cite{bardeen1968non,Ayon-Beato:2000mjt}, regular black holes constructed by Dymnikova \cite{dy92,Dymnikova:2001fb}, the Hayward black hole \cite{hayward}, regular black holes constructed within the gravitational asymptotic safety program \cite{Bonanno:2000ep}, and Planck stars motivated from loop quantum gravity \cite{Rovelli:2014cta,Saueressig:2015xua}. For the sake of conciseness, we will limit the discussion to the Hayward (H) geometry and the regular black holes found in the gravitational asymptotic safety program (RG-improved black holes). 

\begin{figure}[t]
	\sidecaption[t]
	\includegraphics[width = 7cm]{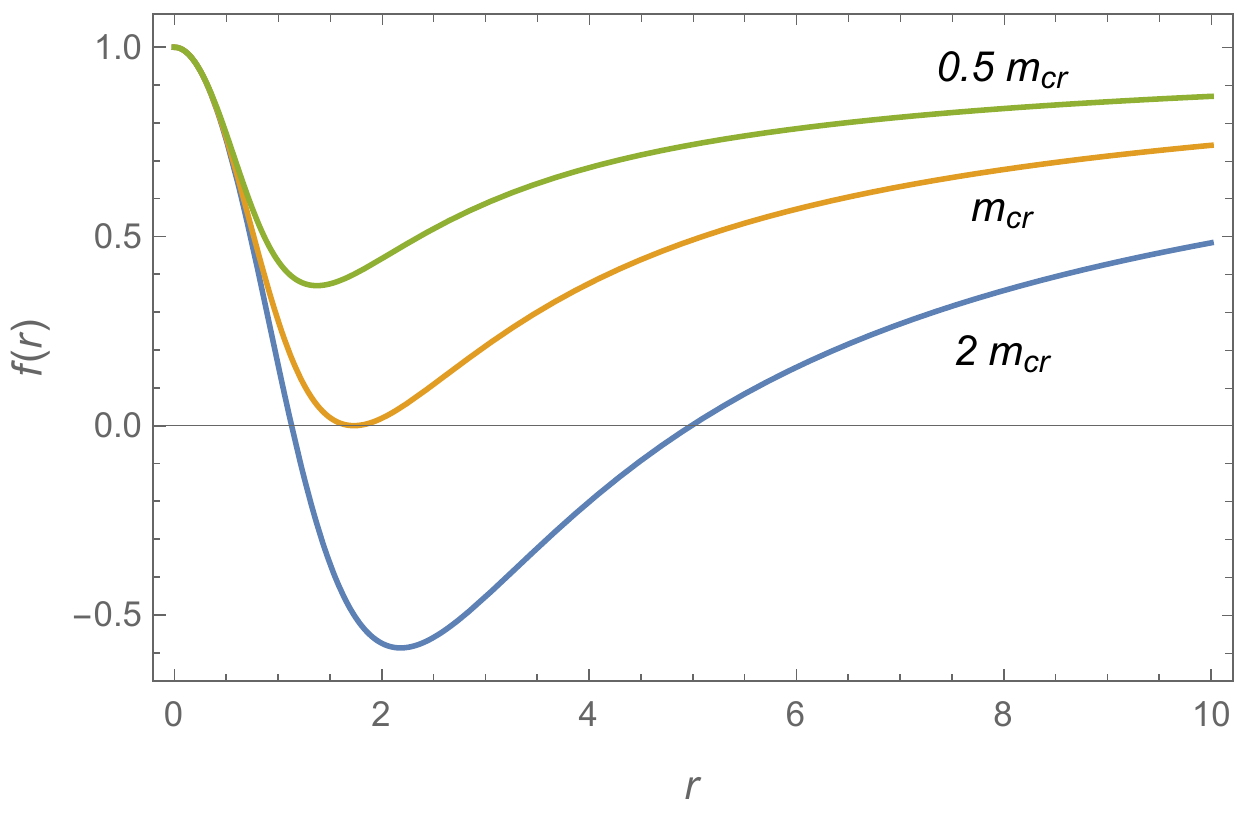}
	\caption{Illustration of the horizon structure of the Hayward black hole \eqref{ma}. For $m > m_{cr}$ one encounters an outer event horizon and an inner Cauchy horizon. These horizons merge for $m=m_{cr}$. For $m < m_{cr}$ no horizons appear. The graph has been obtained for a Hayward geometry with $l=1$.}
	\label{fig:1}      
\end{figure}
\medskip
\emph{Hayward black holes.} Perhaps the simplest implementation of the above requirements has been proposed by Hayward 
in \cite{hayward}. In this case, the function $f$ is given by 
\begin{equation}
	\label{ma}
	f=1-\frac{2 M(r)}{r},\quad\quad    M(r) =\frac{m r^3}{r^3+2 m l^2} \, . 
\end{equation}
Here $l$ is a characteristic scale of the order of Planck length. Interestingly, (\ref{ma}) was actually first derived by Poisson and Israel \cite{Poisson:1988wc} assuming a simple relation between vacuum energy and curvature via the effective Einstein field equations (\ref{efe}). In the ``evolutionary" picture of the interior described in the introduction one would then expect that a consistent theory of quantum gravity should be able reproduce a similar behavior at smaller radii.

Following the general discussion of Sect.\ \ref{sect.2.1}, the explicit form of the horizon condition \eqref{horizon} for the static Hayward geometry is
\be\label{Hay.horizon}
2 l^2 m - 2 m r^2 + r^3 = 0 \, . 
\ee
For $m > m_{\rm cr}$ this equation has two real, positive roots $r_+ > r_-$, i.e., one encounters an event and one Cauchy horizon. For the critical mass $m_{\rm cr} = \frac{3\sqrt{3}}{4} l$, the two horizons coincide and the black hole is extremal. For $m < m_{\rm cr}$ one has a regular geometry without horizons. This characteristic structure of $f(r)$ is illustrated in Fig.\ \ref{fig:1}. It is universal in the sense that it is essentially identical for all regular black holes building on a de Sitter core. 

\begin{figure}[t!]
	\includegraphics[width = 0.48\textwidth]{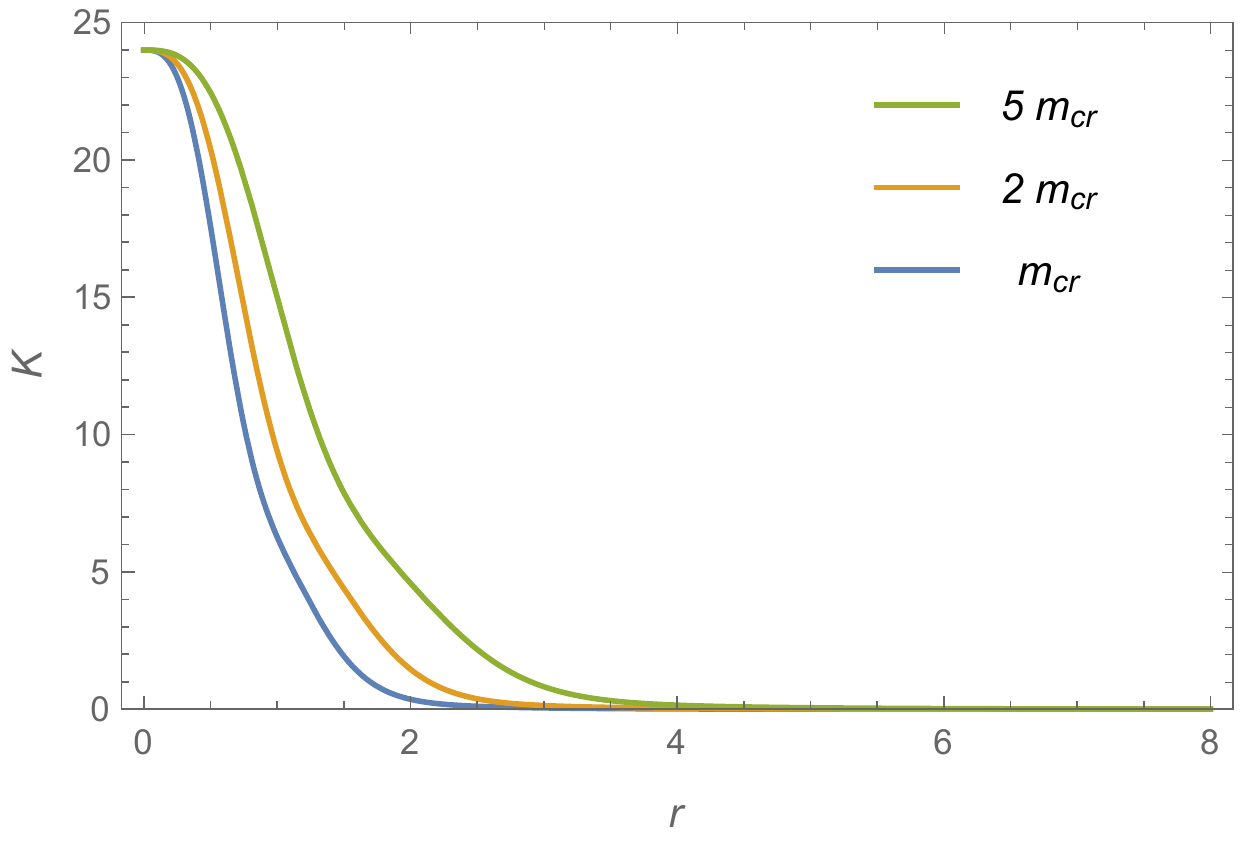} \, 
	\includegraphics[width = 0.48\textwidth]{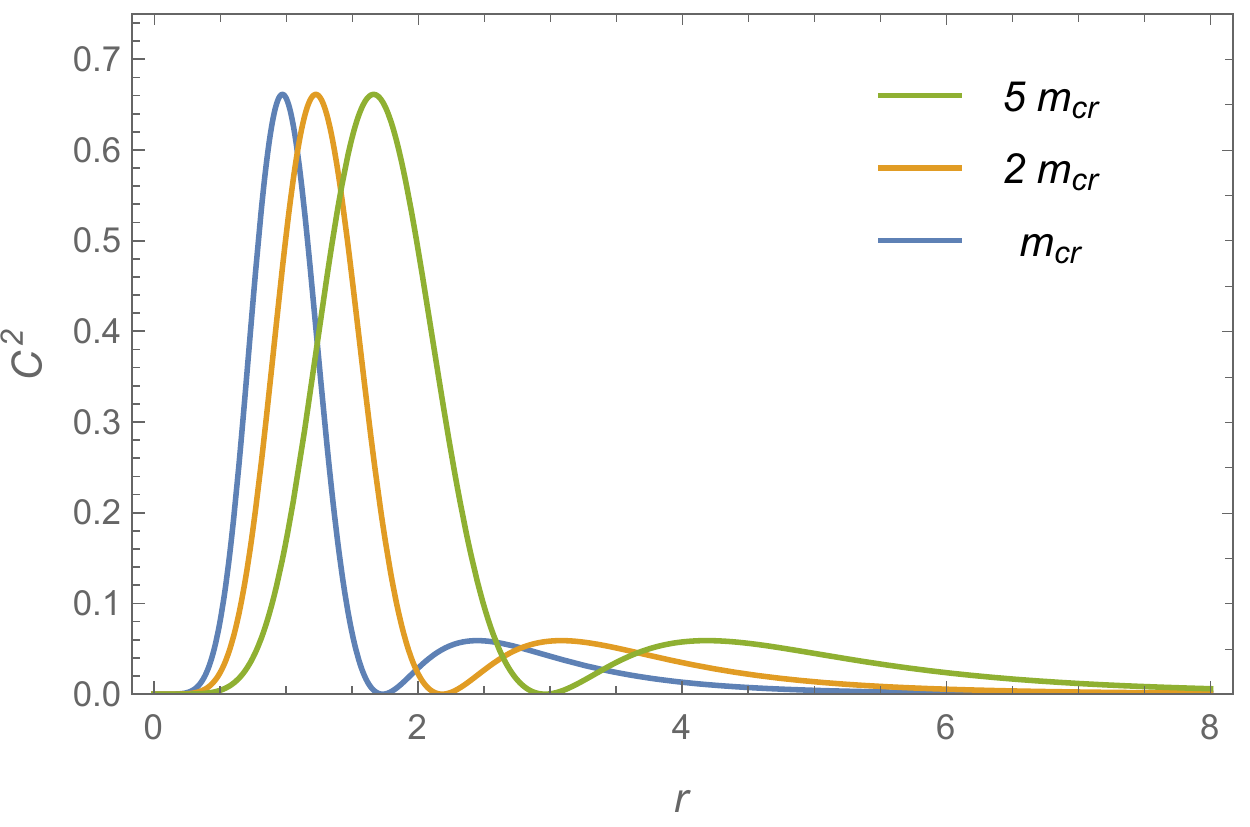}
	\caption{Illustration of the Kretschmann scalar $K$ (left) and the squared Weyl tensor $C^2$ (right) evaluated for the Hayward geometry \eqref{ma} with $m= m_{\rm cr}$ (blue line), $m=2m_{\rm cr}$ (orange line), and $m = 5 m_{\rm cr}$ (green line). The curvature invariants are finite everywhere and the geometry obeys the limiting curvature hypothesis. The graphs have been obtained for $l=1$.}
	\label{fig:2}      
\end{figure}

It is also straightforward to evaluate the curvature scalars \eqref{eq.curvature} for the Misner-Sharp mass \eqref{ma}. The result is shown in Fig.\ \ref{fig:2} for masses $m$ given by multiples of $m_{\rm cr}$. The Kretschmann scalar reaches its maximum at $r=0$ and decreases monotonically with increasing $r$. The square of the Weyl tensor exhibits zeros at the origin and one specific point located between the horizons $r_-$ and $r_+$. Thus, the regular \emph{static} black hole geometry is compatible with the limiting curvature hypothesis \cite{markov1982ultimate,frolov1989through,frolov1990black} for all values $m$.

\medskip
\emph{RG-improved black holes.} Interestingly, the gravitational asymptotic safety program \cite{Percacci:2017fkn,Reuter:2019byg} has provided non-trivial hints that the theory supports regular black holes with a de Sitter core. Ref.\ \cite{Bonanno:2000ep} applied the method of renormalization group (RG)-improvement to a Schwarzschild black hole in order to obtain an effective geometry taking quantum gravity corrections into account. This leads to a Misner sharp mass exhibiting singularity resolution. In this case, the effective running of the Newton constant at high energies produces an effective mass $M(r)$ which vanishes as $r^3$ at small distances. The essential elements of the construction, leading to the geometry \eqref{RGBH}, can be summarized as follows.

Investigating asymptotic safety based on solutions of the Wetterich equation adapted to gravity \cite{Reuter:1996cp} leads to a scale-dependent Newton's coupling $G(k)$ depending on a coarse graining scale $k$. Using the arguably simplest approximate solution of the Wetterich equation based on the Einstein-Hilbert truncation \cite{Reuter:1996cp,Lauscher:2001ya,Reuter:2001ag}, this scale-dependence can be approximated by \cite{Bonanno:2000ep}
\be\label{2.23}
G(k)=\frac{G}{1+\omega \; G \; k^2} \, . 
\ee
Here $G$ is the laboratory value of Newton's coupling measured at $k=0$ and $\omega$ is a \emph{positive} constant.  At large distances, 
$k\rightarrow 0$, $G(k)$ approaches $G$. At short distances, $k\rightarrow \infty$, the scaling $\lim_{k \rightarrow \infty} k^2 G(k) = \omega^{-1} = g_* > 0$ is dictated by the Reuter fixed point providing the UV-completion of the theory.

Following the original derivation \cite{Bonanno:2000ep}, the RG-improvement process starts from the function $f(r)$ describing a classical Schwarzschild black hole. Subsequently $G \mapsto G(k)$ is promoted to the scale-dependent coupling \eqref{2.23}. The RG-improved geometry is then obtained by identifying the coarse-graining scale $k$ with the inverse radial proper distance $d(r)$ between the origin and a point located at radius $r$, $k^2 = \xi^2/d(r)^2$. This procedure leads to the geometry \cite{Bonanno:2000ep}
\begin{equation}\label{RGBH}
	f(r) = 1 - \frac{2 G m r^2}{r^3+ {\widetilde\omega} G (r + \gamma m G)} \, , \qquad   M(r) = \frac{m r^3}{r^3+ {\widetilde\omega} G (r + \gamma m G)} \, . 
\end{equation}
Here $\widetilde{\omega}\equiv \omega\xi^2$ is a positive constant and $\gamma\approx 9/2$ but the qualitative features of the model are independent of its precise value. We now return to geometric units $G=1$. Moreover, we will set $\gamma=0$. While this does not strictly correspond to a regular black hole with a de Sitter core in the sense of eq.\ \eqref{eq.regularityM}. Nevertheless, this limit still captures all essential features of the horizons exhibited by the regular model while significantly simplifying the discussion. The case $\gamma =9/2$ can be found in \cite{Bonanno:2000ep}.

Qualitatively, the features of the function $f(r)$ describing the RG-improved black hole are identical to the ones shown in Fig.\ \ref{fig:1}. The horizon condition resulting from \eqref{RGBH} is $(\gamma=0)$
\be\label{RGhorizon}
r \, \left( r^2 - 2 m r + \widetilde{\omega} \right) = 0 \, . 
\ee
Analogous to the Hayward case, there exists a critical mass value
\begin{equation}
	\label{sei}
	m_{\rm cr}=\sqrt{\widetilde{\omega}} \, . 
\end{equation}
For $m>m_{\rm cr}$,  $f(r)$ has two simple zeros at
\be\label{sette}
r_{\pm} =  m \pm \sqrt{m^2-m_{\rm cr}^2} \, . 
\ee
Hence, the spacetime has an outer event horizon at $r_{+}$ and an inner (Cauchy) horizon at $r_{-}$. For $m=m_{\rm cr}$ the two horizons coincide and there is one double zero at $r_{+}=r_{-} = \sqrt{\widetilde{\omega}}$ and the black hole is extremal.  If $m<m_{\rm cr}$, the spacetime is free from any horizon.

\subsection{Hawking effect and final state}
\label{sect.33}
So far, our exposition has been limited to static geometries. Following Hawking's seminal work \cite{Hawking:1975vcx}, it is expected that black holes emit thermal radiation in form of a Hawking flux. This radiation comes with a perfect black body spectrum with temperature
\be\label{BHT}
T_{\rm BH} = \frac{\kappa_+}{2\pi} \, , 
\ee
where $\kappa_+$ is the surface gravity at the event horizon \eqref{surface-grav}. The resulting energy loss leads to a decrease of the black hole's mass. For a Schwarzschild black hole with just an event horizon this entails that the black hole evaporates completely within a finite proper time interval. This can be traced back to the temperature being inversely proportional to the mass $m$, so that the black hole turns hotter the lighter it gets.

\begin{figure}[t!]
	\includegraphics[width = 0.48\textwidth]{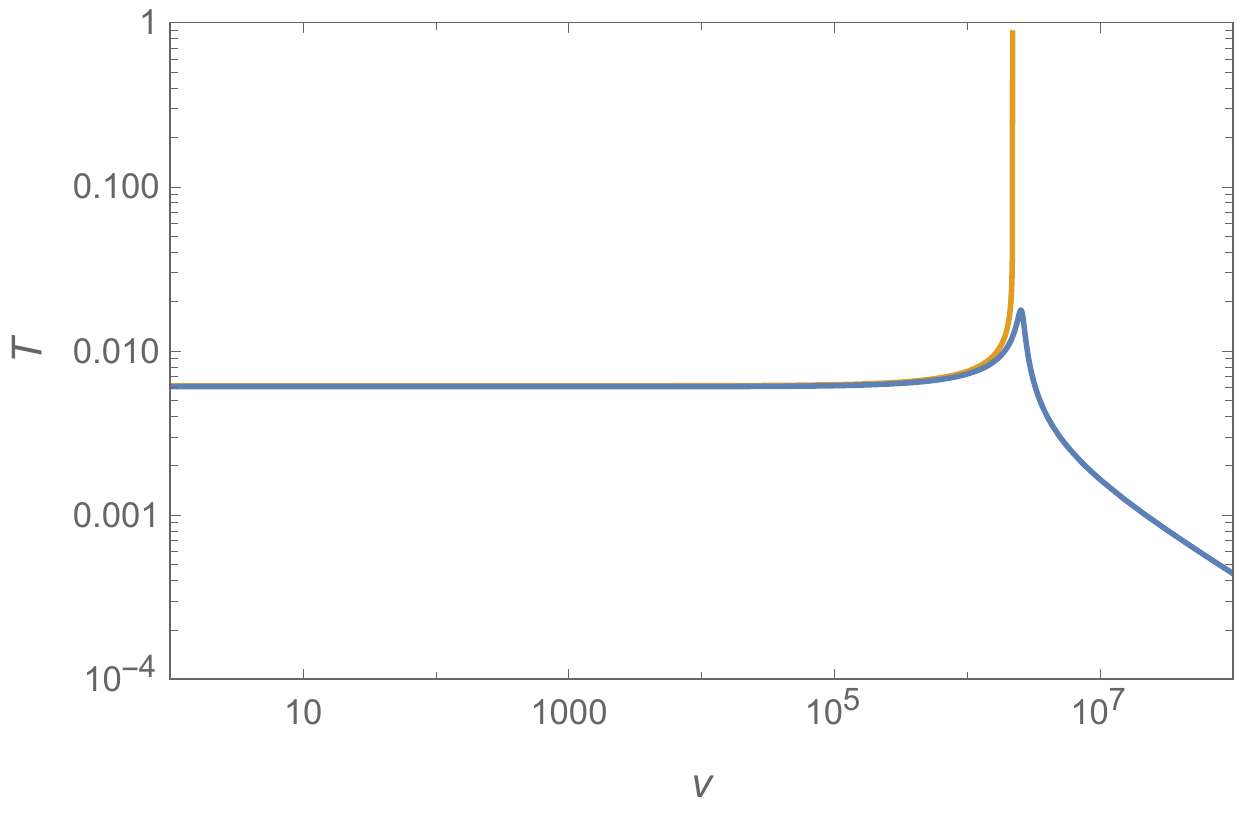} \, 
	\includegraphics[width = 0.48\textwidth]{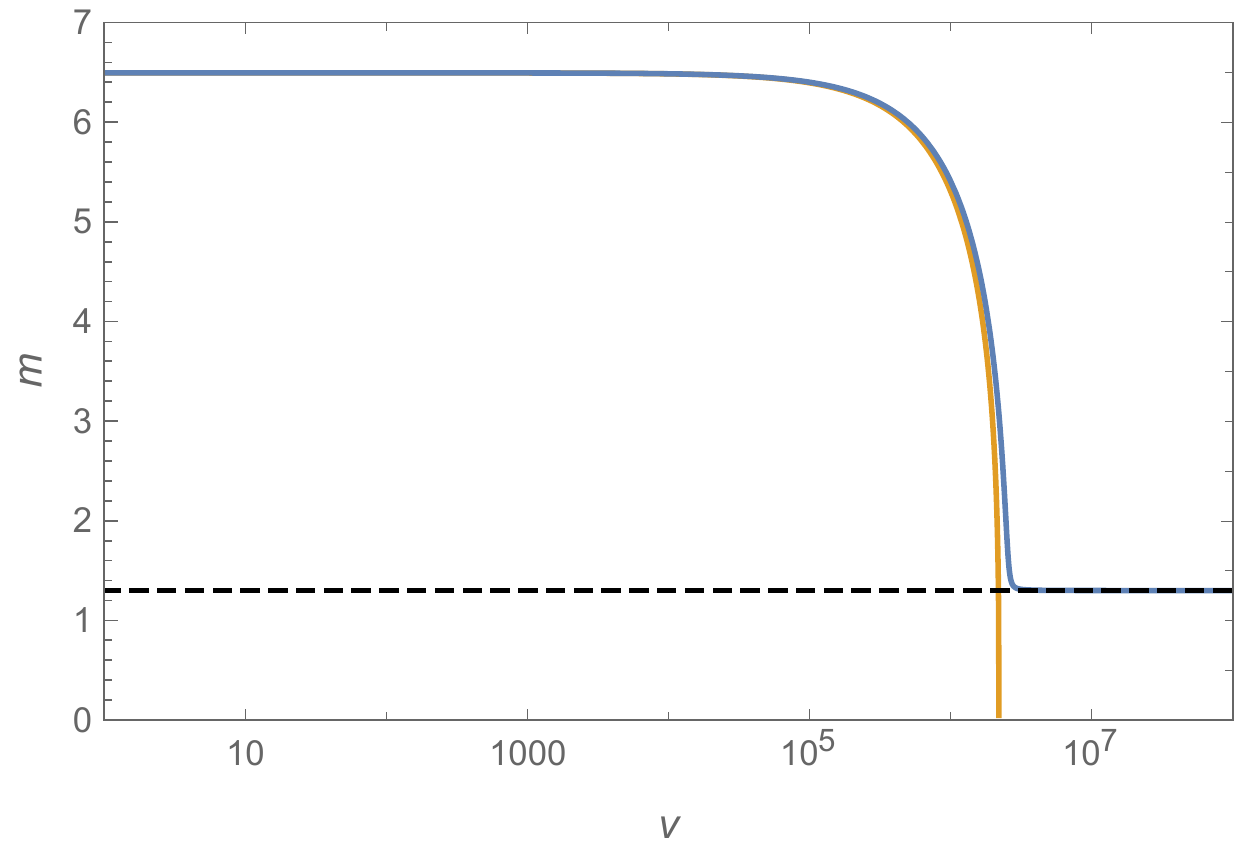}
	\caption{Illustration of the thermodynamical properties of a regular black hole. Its temperature (left) remains finite and the mass (right) approaches the critical mass $m_{cr}$ (dashed line) asymptotically. The graphs have been obtained for a Hayward geometry with $l=1$. The Schwarzschild geometry ($l=0$) has been added as the orange line for comparison.}
	\label{fig:3}       
\end{figure}

The presence of a Cauchy horizon changes this picture drastically. While the black hole is still expected to experience mass loss due to the Hawking flux, the temperature of the radiation remains finite since the two horizons approach each other as the black hole becomes lighter, c.f.\ Fig.\ \ref{fig:1}. As a consequence, the Hawking temperature decreases in the final stage of the evaporation process and the geometry asymptotes to a cold remnant given by the extremal black hole. Figure \ref{fig:3} contrasts these two situations. The black hole evaporation of a Schwarzschild black hole is shown as orange lines indicating that the process terminates in a finite time-span. The generic situation encountered for regular black holes with a Cauchy horizon is shown by the blue curves, illustrating that one obtains a remnant with mass $m_{\rm cr}$.

We illustrate these properties using the technically simplest setting of the RG-improved geometry \eqref{RGBH} with $\gamma=0$. The analysis of other regular black hole geometries featuring a Cauchy horizon follows along the same lines but results in expressions which are significantly more complex. Thus we opt for this example for pedagogical reasons. In order to stress the universal features of the construction, we replace the model parameter $\widetilde \omega$ by $m_{\rm cr}$ using \eqref{sei}.

Starting from \eqref{RGBH} and evaluating \eqref{BHT} gives the black hole temperature as a function of its mass
\be\label{nove}
T_{\rm BH}(m)  = {1\over 4\pi  m}\;{\sqrt{1-\Omega^2}\over 1 +\sqrt{1-\Omega^2}} \, , \qquad \Omega \equiv \frac{m_{cr}}{m} \, . 
\ee
The temperature vanishes for $m \searrow m_{\rm cr}$, 
{\it i.e.}, $\Omega \nearrow 1$. This feature underlies the interpretation of an extremal black hole as a ``cold" remnant.

Given the temperature \eqref{nove}, the luminosity of the black hole can be estimated via the Stefan-Boltzmann law $L= \sigma T^4$, where $\sigma = \pi^2/60$ for a single, massless degree of freedom. The emitted power is obtained by multiplying $L$ with the area of the event horizon $A \equiv 4 \pi r_+^2$,
\be
P(m) = \sigma \left( 4 \pi r_+^2 \right) T^4 \, . 
\ee
Here $T$ and $r_+$ are understood as functions of $m$.
The mass-loss is then determined by solving the mass-loss formula
\be\label{massloss}
\dot{m} = - P(m) \, . 
\ee
For the concrete example of the RG-improved black hole
\be
P(m) = \frac{\sigma}{(4\pi)^3 \, m^2} \, \frac{  \left(1-\Omega ^2\right)^2}{ \big(1+\sqrt{1-\Omega ^2}\big)^2} \, . 
\ee

The final part of the evaporation process is described by those terms in the above expressions which are dominant for $m\searrow m_{\rm cr}$ ($\Omega \nearrow 1$). Expanding \eqref{nove} gives the asymptotic form of the temperature
\be
T(m) \simeq \frac{\sqrt{m-m_{\rm cr}}}{2 \sqrt{2} \pi  \, m_{\rm cr}^{3/2}} \, , 
\ee
and the emitted power
\be
P(m) \simeq \frac{\sigma}{16 \pi^3} \, \frac{  (m-m_{\rm cr})^2}{ m_{\rm cr}^4} \, . 
\ee
Integrating \eqref{massloss} in the asymptotic regime then yields
\be\label{duenove}
m(v) \simeq m_{\rm cr}+\frac{m_0 -m_{\rm cr}}{1+\alpha (m_0-m_{\rm cr})(v-v_0)}
\ee
Here $\alpha\equiv \sigma/(16\pi^3\widetilde{\omega}^2)$, and $v_0$ is a time, already in the 
late-time regime, where $m(v_0)=m_0$ is imposed.
The key result is that for $v\rightarrow \infty$
\begin{equation}\label{eq.evap}
	m(v) - m_{\rm cr} \propto 1/v \, . 
\end{equation}
This result is {\it universal}, i.e. it does not depend on the details of the geometry of the regular black hole \cite{Bonanno:2022jjp}. Eq.\ \eqref{eq.h1} shows that it carries over to the Reissner-Nordstr{\"o}m geometry and the Hayward regular black hole as well. The profound consequence of \eqref{eq.evap} arises from the comparison with Price's law 
\be\label{eq.O10}
m_-(v) \simeq m_0 - \frac{\beta}{(v/v_0)^{p-1}} \, , \qquad p \ge 12 \, ,
\ee
where $m_0$ is the asymptotic mass of the black hole, $\beta > 0$ is a quantity with the dimension of a mass, and $v_0$ is the initial time which we set to one in the sequel.
This shows that the Hawking flux dominates over the gravitational waves contribution which vanishes as $1/v^p$. Hence the Hawking effect may play an important role for the stability of a regular black hole at asymptotically late times.

In order to judge the phenomenological viability of regular black holes, understanding whether these geometries are robust once perturbations are included is highly relevant. This applies in particular to perturbations of the asymptotically extreme geometry with respect to the combined influx of  Hawking radiation and  outflux of the star. At this point we have all the prerequisites to review the current understanding of this situation in the next sections.

\section{The elementary mechanics of mass-inflation}
\label{sect.2}
We start by reviewing the basic mechanisms underlying the mass-inflation effect. Our discussion focuses on the historical analysis carried out in the context of the spherically symmetric Reissner-Nordstr{\"o}m geometry. The key result of the analysis is that the mass function $m_+(v)$ in the vicinity of the Cauchy horizon grows exponentially once perturbations which naturally appear in a gravitational collapse of a star into a black hole are included,
\begin{equation}
	\label{mi}
	m_+(v) \sim v^{-(p-1)} e^{\kappa_{-} v} \, , \qquad \text{as} \quad v \rightarrow \infty \, , 
\end{equation}
with $p$ being the exponent appearing in Price's law \eqref{eq.O10}. We start by giving a qualitative discussion of the mass-inflation mechanism in Sect.\ \ref{sect.2.2}. Sect.\  \ref{sect.22} then derives \eqref{mi} based on 
the two-dimensional wave equation for the generalized mass function in the presence of a cross-flow of outgoing and ingoing streams of radiation. We close with a brief discussion on the strength of the resulting signularity in Sect.\ \ref{sect.sing}.
\subsection{Colliding mass-shells and DTR-relations}
\label{sect.2.2}
\begin{figure}[t!]
	\begin{center}
		\includegraphics[width = 0.48\textwidth]{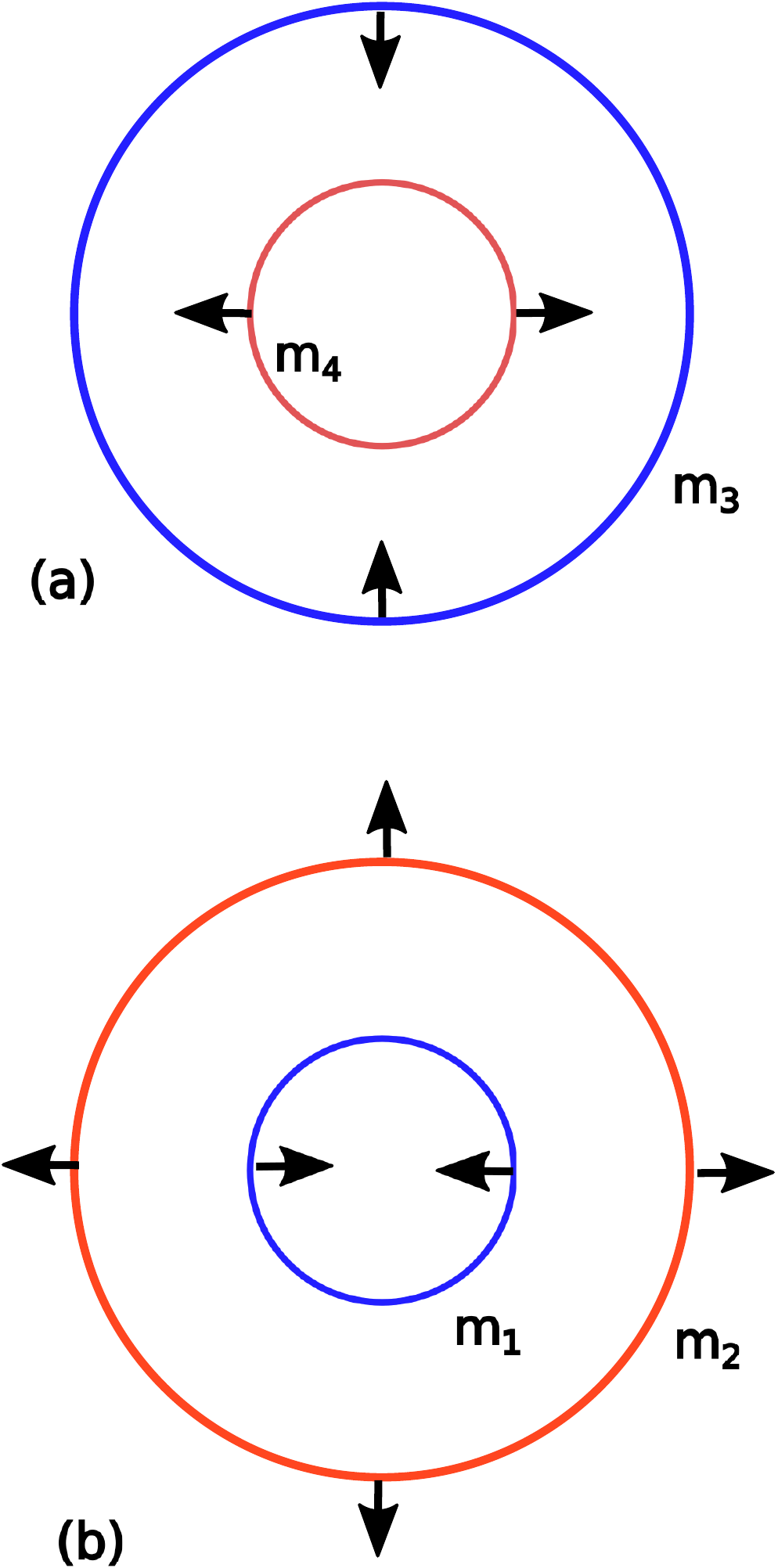} 
	\end{center}
	\caption{Two spherically symmetric, transparent, concentric shells colliding at the speed of light: 
		(a) before collision, (b) after the collision. 
		The blue circle represent ingoing radiation close to the Cauchy horizon, while the red shell
		represents the outgoing radiation.}
	\label{fig:shell1}      
\end{figure}
Qualitatively, the divergence of the local mass in the interior of a realistic black hole can be understood with the following example (see \cite{Hu:1995jr}). Let us consider a time-like moving shell of radius $R(\tau)$ where $\tau$ is the proper time. The shell divides spacetime into a region $\cM_+$ inside and $\cM_-$ outside the shell and we assume that the mass functions $m_+$ and $m_-$ in each sector are constant. The proper mass of the shell, $m_{\rm shell}$, satisfies %
\begin{equation}
	dm_{\rm shell} + P \, d(4\pi R^2)=0
\end{equation}which implies mass conservation if $P=0$.  The equation of motion of the shell  is ruled by  \cite{1970NCimB..67..136C} 
\begin{equation}
	m_{-}-m_{+} = m_{\rm shell} \left ( 1- \frac{2 m_{+}}{R}+\left (\frac{ dR}{d\tau} \right) ^2\right)^{1/2}-\frac{m_{\rm shell}^2}{2 R} 
\end{equation}It expresses the total conserved gravitating mass $m_{-}-m_{+}$ of the shell as a sum of four terms (expanding the square root to first order): the rest-mass $m_{\rm shell}$, the kinetic energy $\frac{1}{2} m_{\rm shell} \dot{R}^2$, the mutual potential energy $-m_{\rm shell} m_{+}/R$, and a self-potential energy $-\frac{1}{2}m_{\rm shell}^2/R$. Note that the ``potential energy'' contributes to the total mass of the {\it outer} body, it is a ``binding energy'' \cite{Blau:1989zs}. If the outer body can be released from its gravitational binding, its gravitating mass can increase. 

This can be easily illustrated in the light-like limit. In this case, we may gain intuition from the  DTR-relations, first discovered by Dray, 't Hooft, and Redmount \cite{Dray:1985yt,1985PThPh..73.1401R}, and subsequently generalized in \cite{bar91}. 
In the latter work the (generalized) DTR-relations are derived as geometrical consistency conditions on the function $f$, beyond spherical symmetry,
for an energy-momentum tensor which is vacuum except for a distributional delta function source representing the shells, also see \cite{Blau:1989zs} for an instructive discussion.

The idea behind the DTR-relations is to consider the collision of two infinitely thin, light-like, pressureless shells describing an ingoing and outgoing perturbation, see Fig.\ \ref{fig:shell1}.\footnote{For more details on the infinitely thin-shell formalism, see \cite{Israel:1966rt}.}  The shells provide a highly idealized model of matter perturbations in the black hole interior which are naturally expected from the collapse of a star into a black hole. They capture the essential features underlying the mass-inflation effect, the counter-streaming of matter located between the inner and outer horizon of the black hole in the optical geometric limit \cite{Hamilton:2008zz}. 

\begin{figure}[t!]
	\begin{center}
		\includegraphics[width = 0.48\textwidth]{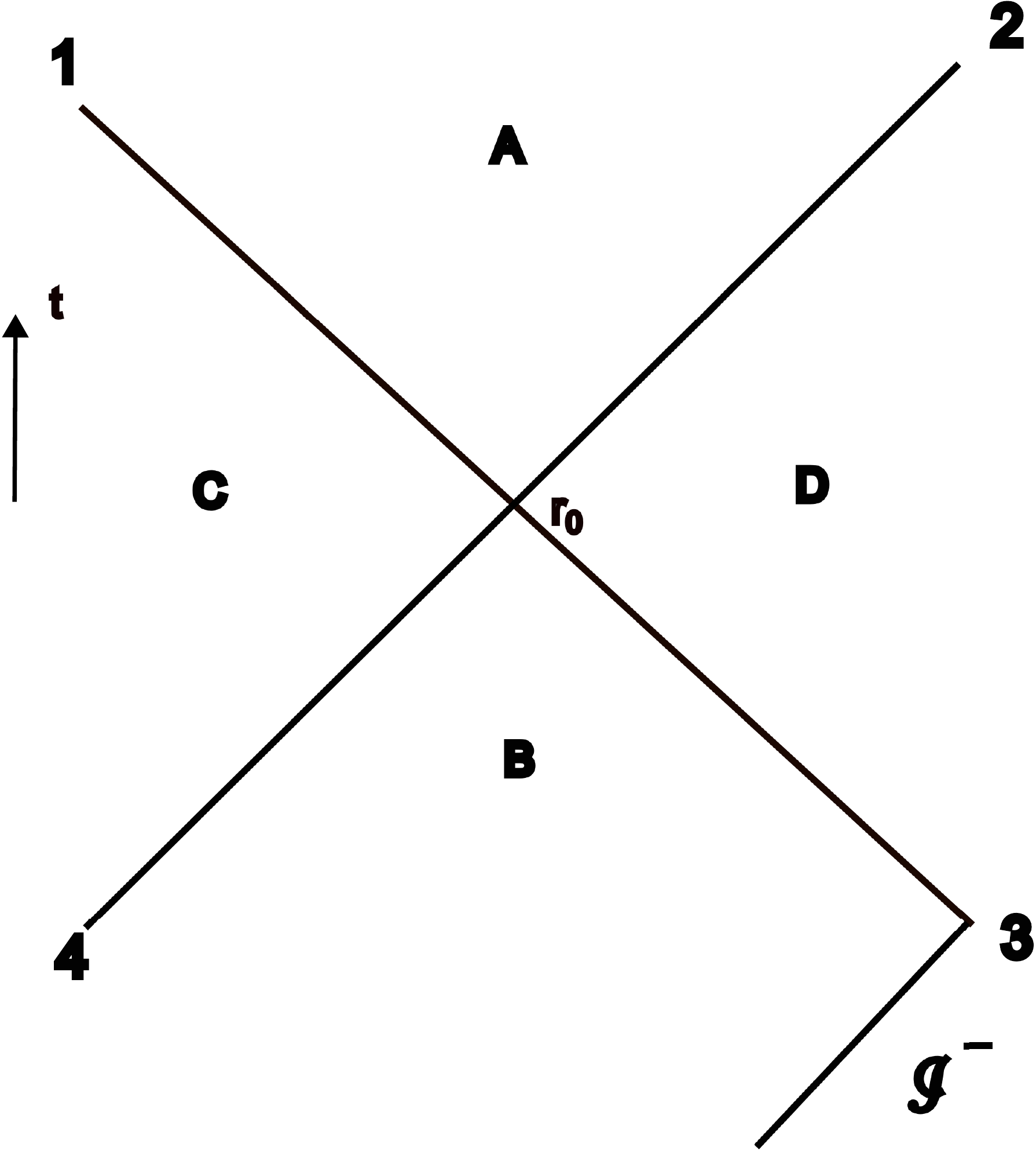} 
	\end{center}
	\caption{Fig.\ \ref{fig:shell1} recast as a spacetime diagram: $r_0$ is the colliding radius
		and sectors $B$ and $C$ represent the future and past evolution of the spacetime, respectively. }
	\label{fig:shell2}      
\end{figure}

As depicted in Fig.\ \ref{fig:shell2}, the ingoing and outgoing shells separate spacetime in four regions $\cM_A, \cM_B, \cM_C, \cM_D$. The generalized DTR-relations then encode the fact that the spacetime metric in each region must agree at the collision point $r_0$. We then assume that in each region the metric takes the form \eqref{eq:eddfink} with $f_i(v,r) = f_i(r)$, $i=A,B,C,D$, being static. Following the pedagogical derivation given in \cite{Brown:2011tv}, this entails that 
\be\label{eq:DTR1}
f_A \, f_B = f_C \, f_D \, , \quad \text{at} \quad r_0 \, . 
\ee
For the sake of simplicity, let us assume that the metric in each sector has the form of the Schwarzschild metric. Assuming that the region $\cM_C$ corresponds to flat space, the functions $f_i$ in each region take the form
\be\label{feq1}
f_C = 1 \, , \qquad f_A = 1 - \frac{2m_1}{r_0} \, , \qquad f_B = 1 - \frac{2m_4}{r_0} \, , 
\ee
together with
\be\label{feq2}
f_D = 1 - \frac{2 (m_3+m_4)}{r_0} =  1 - \frac{2 (m_1+m_2)}{r_0} \, . 
\ee
The relation \eqref{feq2} implies the conservation law
\begin{equation}
	m_1+m_2=m_3+m_4 \, . 
\end{equation}
Moreover, evaluating \eqref{eq:DTR1} gives the additional conditions
\begin{equation}
	\label{d1}
	m_1=m_3(1-2 m_4/r_0)^{-1},\quad m_2 = m_4 (1-2m_1/r_0) \, . 
\end{equation}

We now consider a collision just outside the horizon of the interior field, $r_0 = 2 m_4 + \delta$
with $\delta\rightarrow 0^+$. The first identity in \eqref{d1} then entails that, close to the inner horizon, the mass increase $m_1-m_3$ diverges
as 
\begin{equation}
	m_1-m_3 = \frac{2 m_3 m_4}{\delta} \, . 
\end{equation}
This divergence implies that the new Schwarz\-schild mass of the outgoing shell would then become negative because of its potential energy. 

This example, albeit very simple,  illustrates the basic physical mechanism underling the mass-inflation phenomenon: the imploding shell can mimic the fallout from the radiative tail of the collapse while the outgoing shell models the outflow from the collapsing star as depicted
in Fig.\ \ref{fig:shell1}. In a spacetime diagram the Cauchy horizon would develop
beyond the history of region A, extending region B and the collision point $r_0$ is 
near the actual Cauchy horizon. The fact that the DTR-relations are not limited to spherical symmetry then suggests that the same phenomenon also appears in less symmetric situations including a Kerr black hole \cite{1991PhLA..161..223B}.
%

This is the best that can be concluded on the basis of DTR-relations. In particular, promoting $m_1 - m_3$ to a function of $v$ is beyond the framework and should not be used to draw conclusions on the strength of the singularity, albeit this point is sometimes not appreciated in the literature \cite{Carballo-Rubio:2018pmi,2021PDU....3300853B}.
Determining the strength of the divergence requires a more detailed analysis based on dynamical models. We cover this in the next subsection.
\subsection{Dynamical models of mass inflation}
\label{sect.22}
The simple analogy discussed in the previous section can be useful to understand the interior of realistic black holes. In this case, it is useful to make some simplifying assumptions.  A realistic,  rotating (non-spherical) black hole may be schematized as a spherically symmetric, charged black hole because their horizon structures are similar. Moreover, the tail of gravitational quadrupolar waves may be idealized as spherical waves. Since the perturbations get blue-shifted  near the Cauchy horizon,  one can use an ``optical'' approximation and describe the matter infalling onto the Cauchy horizon as a stream of light-like particles. 

In 1981, Hiscock \cite{1981PhLA...83..110H} treated the dynamics of this situation by considering the charged Vaidya metric,
\begin{equation}
	ds^2 = 2 dr dv - \left (1- \frac{2 m(v)}{r}+\frac{Q^2}{r^2} \right ) dv^2 + r^2 d\Omega^2 \, , 
\end{equation}
promoting the mass function of the Reissner-Nordstr{\"o}m metric to a $v$-dependent function. Then $m(v)$ is the externally measured mass of the black hole which varies with advanced time $v$ because of the inflow. The associated stress-energy tensor (the electrostatic component is irrelevant in this discussion) corresponds to radially infalling light-like dust:
\begin{equation}
	\label{stress}
	T_{\alpha\beta}=\frac{\dot{m}(v)}{4 \pi r^2}l_\alpha l_\beta, \quad l_\alpha = -\partial_\alpha v \, , \quad l_\alpha l^\alpha=0 \, . 
\end{equation}
The ansatz
\begin{equation}
	\label{tail}
	m(v) = m_0 - \frac{\beta}{(v/v_0)^{p-1}}
\end{equation}
reproduces the power-law decay of the radiative tail. (For the generic case of the ``quadrupole'' waves $p=12$.)
It is easy to show from (\ref{stress}) and (\ref{tail}) that an observer approaching the Cauchy horizon ($v\rightarrow \infty$)  reaches the horizon in a finite proper time and measures an energy flux diverging like $e^{2\kappa_{-} v}$, with $\kappa_{-}$ being the surface gravity \eqref{RN.surfacegrav} for the asymptotic, stationary Reissner-Nordstr{\"o}m black hole of mass $m_0$ and charge $Q$.

In \cite{Poisson:1990eh}, Poisson and Israel extended this model by including a crossflow of radially ingoing and outgoing radiation. This analysis shows that the geometry near the Cauchy horizon changes dramatically. Approaching the Cauchy horizon, the Coulomb component of the Weyl curvature
\be\label{psi2ev}
\Psi_2 = - \frac{m(v)}{2r^3} + \frac{Q^2}{r^4} \, , 
\ee
and hence the mass parameter $m(v)$ diverge like
\begin{equation}
	\label{mi2}
	m(v) \simeq v^{-(p-1)} e^{\kappa_{-} v}, \quad v \rightarrow \infty \, . 
\end{equation}
This behavior has to be contrasted with the situation outside the event horizon, where $m$ approaches its ADM value at infinity. 

It is interesting to see how this effect originates from the point of view of the field equations. A spherically geometry 
can be described by the metric \eqref{eq:lineelement1}.
The Einstein equations can be reformulated as a two-dimensional, covariant wave equation for $m$,
\begin{equation}
	\label{wavem}
	\Box \, 
	m = - 16 \pi^2 r^3 T^{ab}T_{ab} \, . 
\end{equation}
Going to Kruskal coordinates, the energy momentum tensor for a null-crossflowing radiation can 
be written as
\begin{equation}\label{Tabans}
	T_{ab} = \frac{L_{\rm in} (V)}{4\pi r^2} \partial_a V \partial_b V+
	\frac{L_{\rm out} (U)}{4\pi r^2} \partial_a U \partial_b U \, .
\end{equation}
Due to $\p_a V$ and $\p_a U$ being null vectors, the square of the energy momentum tensor vanishes if only one component $T^{\rm out}_{ab}$ or $T^{\rm in}_{ab}$ is considered. In this case the source on the right-hand side of eq.\ \eqref{wavem} vanishes and there is no mass-inflation. It is only when \emph{both components} $T^{\rm out}_{ab}$ or $T^{\rm in}_{ab}$ are present that the wave equation contains a non-trivial source term triggering the growth of $m$. 

At this stage, one has to adopt assumptions for $L_{\rm in}$ and $L_{\rm out}$. $L_{\rm in}$ uses Price's law \eqref{tail}, which in Kruskal coordinates $V$ reads
\begin{equation}
	L_{in}(V) =\frac{\beta}{(-\kappa_{-} V)^2} (-\ln(-\kappa_{-} V))^{-p}. 
\end{equation} 
$L_{\rm out}(U)$ receives two contributions, one from the collapsing star and a second one from gravitational waves reflected by the inner potential barrier. At late advanced time the first one is negligible. The second one is basically following Price's law, but written in terms of the coordinate $U$. Thus, we have control over $L_{\rm out}(U)$ at asymptotically late times only and every analysis going beyond this regime has to track the full dynamics numerically.

Near the Cauchy horizon,\footnote{This makes the crucial assumption that the Cauchy horizon actually exists. While there is good evidence for forming such a horizon for the black holes encountered in general relativity, the question whether such a horizon actually forms in the collapse to a regular black hole is a largely open question, see \cite{Ziprick:2010vb,Biasi:2022ktq} for some recent work in this direction. For a first self-consistent dynamical calculation of mass-inflation without assuming the existence of a Cauchy horizon we refer to \cite{Bonanno:1994qh}.} assuming that $r\approx r_0$, equation (\ref{wavem}) then reads
\begin{equation}
	\partial_{U} \partial_V m= -\frac{2}{r_0} L_{\rm in} (V) L_{\rm out} (U) \,. 
\end{equation}
In the limit $V \rightarrow 0_{-}$ this relation is readily integrated, producing a divergence of the type
\begin{equation}
	\label{mikruskal}
	m\simeq \frac{1}{-\kappa_{-}V (\ln(-\kappa_{-} V))^{(p-1)}}  \, . 
\end{equation}
In terms of the original advanced time coordinate $v$, 
(\ref{mikruskal}) entails the exponential growths of the mass function underlying the mass-infaltion effect (\ref{mi}).
\subsection{Mass-inflation instability and singularities of spacetime}
\label{sect.sing}
At this stage it is interesting to discuss the implications of the growths of the mass function \eqref{mi} for the singularity structure of spacetime. We recall that a singularity for which all algebraic scalars of curvature are finite and yet some component of the curvature diverge in some frame is known as a ``whimper'' singularity. In most of the cases these singularities are unstable and tend to evolve in a strong, ``scalar'' singularity. As demonstrated in the previous section, it turns out that if the Hiscock model is perturbed with an additional  {\it outflow} of radiation, which in a realistic collapse would always be present either for the presence of the collapsing star or for the scattering gravitational waves from the inner potential barrier, the Weyl curvature diverges without limit at the Cauchy horizon. 

The strength of the resulting singularity can then be understood as follows. We focus on the future sector of the shell, denoting the coordinates and mass function in this sector with the subscript ``$+$''. Using the advanced coordinate, the asymptotic form of the metric near the Cauchy horizon reads
\be\label{eq:cauchyasym}
ds^2  \simeq 2 \frac{dv_+}{r} \left( r dr + m_+(v_+) dv_+ \right) + r^2 d\Omega^2 \, . 
\ee
We then define a new coordinate $u$ through the relation $du = (rdr + m_+(v_+)dv_+)$. This coordinate is 
regular at the Cauchy horizon. The line-element \eqref{eq:cauchyasym} then becomes \cite{Bonanno:1994qh}
\be
ds^2 \simeq 2 \frac{dv_+ du}{r} + r^2 d\Omega^2 \, . 
\ee
Again, this expression is manifestly regular at the Cauchy horizon. Since it is possible to find a coordinate system where the metric is regular, the singularity building up at the Cauchy horizon is rather weak. This fact has profound consequences: as already realized by Ori \cite{ori91} and further investigated
by Burko \cite{Burko:1999zv}, the mass-inflation singularity does not satisfy the necessary conditions to be strong in the Tipler sense \cite{Tipler:1977zza}.\footnote{According to Tipler, a null singularity is called ``strong'' if there exists at least one component
	of the Riemann tensor (in a parallelly propagated frame) which does not converge when integrated with respect to the affine parameter $\tau$ twice. The physical meaning of this requirement is that the tidal distortion is not finite as an observer crosses the singularity.}  A measure of the tidal distortion experienced by an observer is obtained by integrating the square of the Weyl curvature twice. In the case of the standard mass-inflation scenario one finds
\be\label{eq:tidal}
(\Psi_2)^2 \simeq C_{\mu\nu\rho\sigma} C^{\mu\nu\rho\sigma} \simeq \frac{1}{(\kappa_- V)^2 (\log(-\kappa_- V))^{2(p-1)}} \, . 
\ee
Here $\Psi_2$ is the Coulomb-component of the Weyl curvature \eqref{psi2ev}  and  $V \propto \tau$ is proportional to the proper time of an observer impacting on the horizon. The tidal
distortion is obtained by twice integrating \eqref{eq:tidal} and is therefore finite. It has further been argued by Ori
that this behavior could be sufficient to determine a $C^1$-extension of the spacetime beyond the Cauchy horizon \cite{ori91}. However,
according to Kr\'{o}lak, eq.\ \eqref{eq:tidal} still signals a strong singularity, as the expansion of the congruence is divergent \cite{1987JMP....28.2685K}: if the components of the Riemann tensor are integrated only once, the integral does not converge on the singularity.

\section{Mass-inflation for regular black holes}
\label{sect.4}
We now embark on the central theme of this chapter, contrasting the mass-inflation scenario for Reissner-Nordstr{\"o}m and regular black hole geometries. In the latter case we use the Hayward geometry as an explicit representative. The results are more general though, since the Reissner-Nordstr{\"o}m and Hayward geometry constitute representatives for the two classes of universal late-time behavior encountered in the literature \cite{Bonanno:2020fgp,Bonanno:2022jjp}. The discussion will be based on the Ori-model introduced in Sect.\ \ref{sect.23}. The case of a static background is covered in Sect.\ \ref{sect.41} while the modifications due to the mass-loss generated by the Hawking effect are highlighted in Sect.\ \ref{sect.42}. In order to contrast the structural differences of the mass-inflation effect for the Reissner-Nordstr{\"o}m case and regular black holes, we discuss the two geometries in parallel. 
\subsection{The Ori-model -- general setup and dynamics}
\label{sect.23}
The Ori-model constitutes a simplification of the original Poisson-Israel model geared towards making the mass-inflation effect accessible by analytic methods. In this case the outgoing energy flux is modelled  by a spherically symmetric, pressureless null shell $\Sigma$ placed between the (apparent) inner and outer horizon of the black hole geometry. In the language of eq.\ \eqref{Tabans} it corresponds to taking $L_{\rm out}(U)$ to be a delta-function. The shell then acts as a catalyst triggering the mass-inflation instability. In this section, we derive the equations capturing the dynamics of the system for a generic, spherically symmetric black hole spacetime exhibiting an event and a Cauchy horizon.

The shell $\Sigma$ modeling the ingoing perturbation, divides spacetime into a region $\cM_+$ inside and $\cM_-$ outside the shell. Denoting the coordinates on $\cM_\pm$ with subscripts $\pm$, the metric in each sector can be written as
\be\label{eq.O1}
ds^2 = -f_{\pm}(r,v_{\pm}) dv_\pm + 2 dr dv_\pm + r^2 d\Omega^2 \, .
\ee
Insisting that both regions lead to the same induced metric on $\Sigma$ yields that the radial coordinate $r$ can be taken the same in both regions. Eq.\ \eqref{eq.O1} already anticipated this result. In general, the relation between $v_+$ and $v_-$ is non-trivial though. Their dependence is fixed by noting that the position of $\Sigma$ in the two coordinate systems is
\be\label{eq.O2}
f_+ \, dv_+ = f_- \, dv_- \, .
\ee
In practice, we use this relation to express all quantities in terms of the coordinate in the outer sector of the shell, setting $v \equiv v_-$. Eqs.\ \eqref{eq.O1} and \eqref{eq.O2} are independent of the dynamics and completely fixed by the geometrical setup of the model.

Our next task is to find the relation between the Misner-Sharp mass $M_\pm$ in the two sectors. Evaluating Einstein's equations in each sector implies 
\be\label{eq.O3}
\frac{\p M}{\p r} = - 4 \pi r^2 T_v{}^v \, , \qquad \frac{\p M}{\p v} = 4 \pi r^2 T_v{}^r \, ,
\ee
combined with $T_{rr} = 0$. We then introduce the null generators of $\Sigma$
\be\label{eq.O4}
s_{\pm}^\mu \equiv \frac{dx^\mu_\pm}{dr} = \left( 2/f_\pm , 1, 0, 0 \right) \, ,
\ee
where it is convenient to use $r$ as a parameter. Continuity of the flux across $\Sigma$ requires
\be\label{eq.O5}
\left[ T_{\mu\nu} \, s^\mu s^\nu \right] = 0 \, 
\ee
where the square brackets indicate the discontinuity of a scalar quantity across the position of the shell. In terms of the lapse- and the mass-functions, eq.\ \eqref{eq.O5} implies
\be\label{eq.O6}
\left. \frac{1}{f_+^2} \frac{\p M_+}{\p v_+} \right|_\Sigma =
\left. \frac{1}{f_-^2} \frac{\p M_-}{\p v_-} \right|_\Sigma \, . 
\ee
We then recast this equation in terms of the coordinate $v$. Using, eq.\ \eqref{eq.O2}, it is convenient to write
\be\label{eq.O7}
\left. \frac{1}{f_+} \frac{\p M_+}{\p v} \right|_\Sigma = F(v) \, , 
\ee
where
\be\label{eq.O8}
F(v) \equiv \left. \frac{1}{f_-} \frac{\p M_-}{\p v} \right|_\Sigma \, . 
\ee
The introduction of $F(v)$ will facilitate the analysis of the late-time dynamics of the model later on.

Finally, we need an equation determining the dynamics of the shell. Exploiting that $\Sigma$ moves light-like, eq.\ \eqref{eq.O1} gives the relation $f_- \, dv_- = 2 dr$. Thus the position of the shell $R(v)$ follows from 
\be\label{eq.O9}
\frac{dR}{dv} = \left. \frac{1}{2} f_- \right|_{\Sigma} \, . 
\ee
Eqs.\ \eqref{eq.O7} and \eqref{eq.O9} then form a coupled dynamical system determining the position of the shell and the Misner-Sharp mass in its interior in terms of the mass-function $m_-$ in the outer sector of the shell. Since the dynamics of $R(v)$ is independent of $M_+$, one can first study the motion of the shell based on \eqref{eq.O9} before substituting the solution into \eqref{eq.O7}.

The Ori-model applied to a static black hole background fixes $m_-$ by imposing the Price's tail behavior \eqref{eq.O10}.
The Price tail governs the decay of a perturbation at asymptotically late times. We stress that the relation \eqref{eq.O10} applies only asymptotically. At intermediate times one expects that the time-dependence of $m_-(v)$ may be significantly more complicated than indicated by the asymptotic relation.
\subsection{The Ori-model on static backgrounds}
\label{sect.41}
The Misner-Sharp mass for the Reissner-Nordstr{\"o}m (RN) and Hayward (H) black hole is
\be\label{eq.s1}
\begin{split}
	{\rm RN:} \qquad & \, M(r) = m - \frac{Q^2}{2r} \, , \\
	{\rm H:} \qquad & \, M(r) = \frac{m r^3}{r^3 + 2 m l^2}  \, . 
\end{split}
\ee
Here $Q$ is the charge of the black hole and $l > 0$ is a parameter with the dimension of mass which ensures the regularity of the Hayward geometry. The static background analysis includes the effect of the perturbation by identifying the mass function $m_-(v)$ in the outer sector of the shell with the Price-tail behavior given in eq.\ \eqref{eq.O10}.

\begin{figure}[t!]
	\includegraphics[width = 0.48\textwidth]{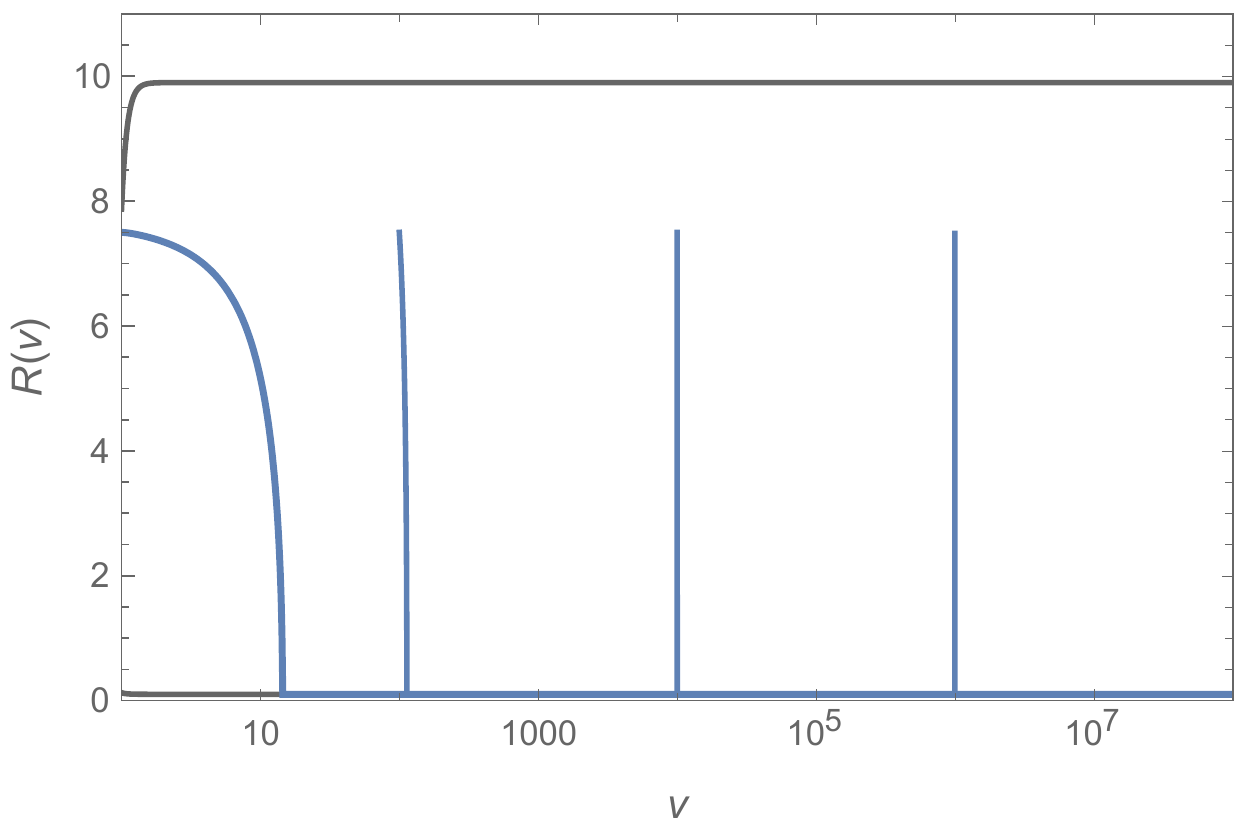} \, 
	\includegraphics[width = 0.48\textwidth]{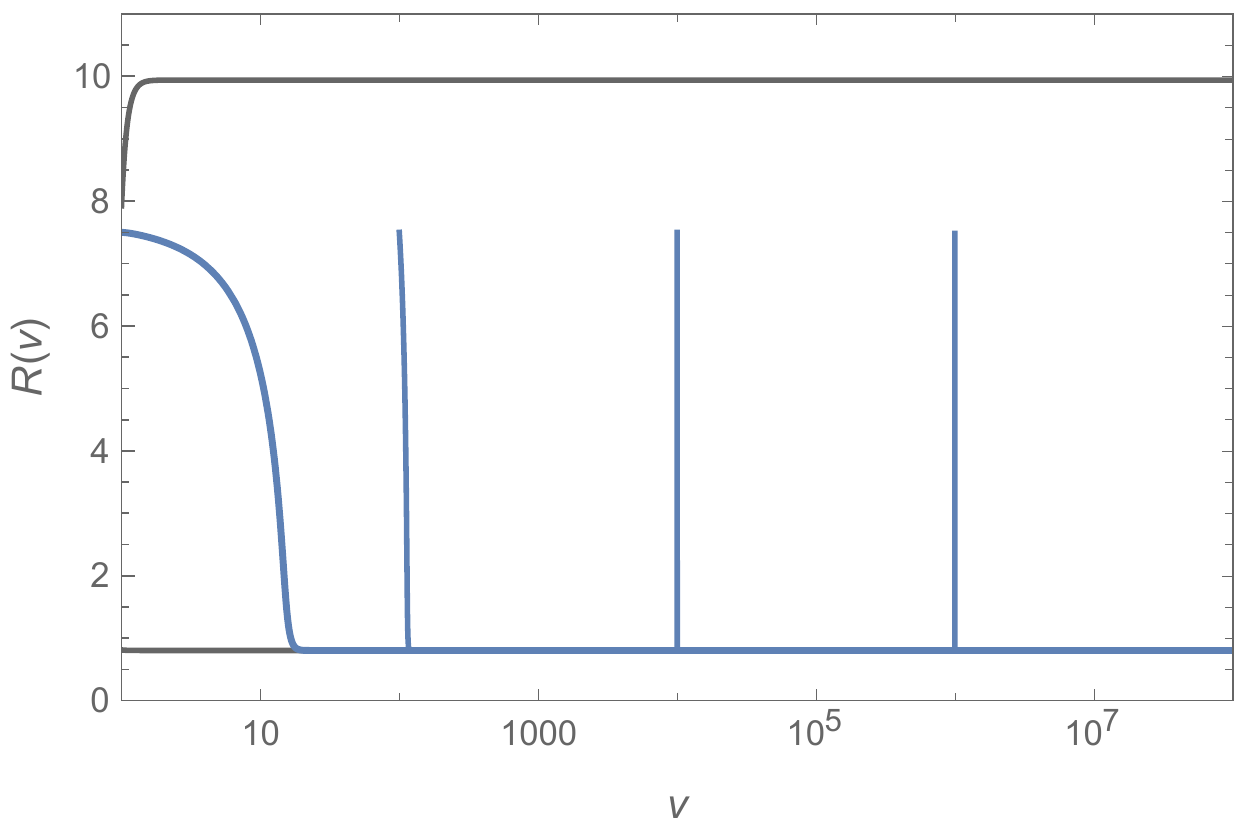}
	\caption{Illustration of the shell dynamics (blue curves) for the background spacetime given by a static Reissner-Nordstr{\"o}m (left) and Hayward black hole (right). The gray lines depict the positions of the apparent event (top) and Cauchy horizons (bottom). The shells reach $r_-$ at asymptotically late times.  The model parameters $Q$ and $l$ have been chosen such that the critical mass of the remnant is $m_{cr} = 1$.}
	\label{fig:s1}    
\end{figure}
Specifying the general equations \eqref{eq.O7} and \eqref{eq.O9} to our exemplary geometries, one finds that the dynamics of the shell is given by
\be\label{eq.s3}
\begin{split}
	{\rm RN}: \qquad \dot{R}(v) = & \, \frac{R^2 -2 m \, R +Q^2}{2 R^2}\, , \\   
	{\rm H}: \qquad  \dot{R}(v) = & \, \frac{R^3 -2 m \, R^2 + 2 l^2 \, m}{2 \left(R^3 + 2 l^2 m \right)}  \, .
\end{split}
\ee
The mass function $m_+(v)$
in the interior is determined from
\be\label{eq.s4}
\dot{m}_+(v) = p(m_+) \, F(v) \, . 
\ee
The explicit forms of the functions $F(v)$ are
\be\label{eq.s5}
\begin{split}
	{\rm RN}: \qquad F(v) = & \, \frac{\dot{m}_-}{R^2 -2 R \,  m_- +Q^2}\, , \\   
	{\rm H}: \qquad F(v) = & \, \frac{\dot{m}_-}{\left( R^3 + 2 l^2 m_- \right) \left(R^3 - 2 (R^2-l^2) m_- \right)} \, , 
\end{split}
\ee
while the dependence of the right-hand side on $m_+$ is captured by the polynomials
\be\label{eq.s6}
\begin{split}
	{\rm RN}: \qquad p(m_+) = & \, R^2 -2 m_+ \, R + Q^2 \, , \\   
	{\rm H}: \qquad p(m_+) = & \, \left( R^3 + 2 l^2 m_+ \right) \left(R^3 - 2 (R^2-l^2) m_+ \right) \, .  
\end{split}
\ee
The polynomials \eqref{eq.s6} encode the key difference of the two geometries: in the Reissner-Nordstr{\"o}m case $p(m_+)$ is linear in the mass while for Hayward it is quadratic. As it turns out, this makes a decisive difference in the stability of the two models.

We proceed by analyzing the late-time dynamics of the two systems. At the analytic level, this is conveniently done by employing the Frobenius method. In the specific case at hand, the $v$-dependent functions are expanded in a generalized power series in $1/v$ which take the general form
\be\label{eq.s7}
f(v) = \frac{1}{v^s} \sum_{k=0}^\infty \, \frac{a_k}{v^{k/2}} \, .
\ee
The parameter $s$ and the coefficients $a_k$ are obtained from substituting this ansatz in the corresponding differential equation, performing an expansion for large $v$, and extracting a hierarchy of equations given by the coefficients appearing at each order in the large-$v$-expansion. This hierarchy is then solved recursively for $s$ (lowest order equation) and the coefficients $a_k$. Note that eq.\ \eqref{eq.s7} already anticipates that consistent solutions require the inclusion of non-integer powers of $1/v$.

Based on this strategy, one finds the following late-time behavior. Starting with the dynamics of the shell, one first establishes that $r_-$ is a fixed point of \eqref{eq.s3} since $f_-|_{r=r_-} = 0$ by definition of the Cauchy horizon. The first correction term obtained from the Frobenius analysis shows that $r_-$ is indeed a late-time attractor,
\be\label{eq.s8}
R(v) \simeq r_- + c v^{-(p-1)} \, , 
\ee
The constant $c$ is a positive coefficient depending on $m_0$ and $\beta$ and again $\simeq$ denotes that the expression holds asymptotically for large $v$. Integrating \eqref{eq.s3} numerically confirms this property. Some sample solutions arising from imposing initial conditions at different values $v$ are shown in Fig.\ \ref{fig:s1}. This confirms that the shell impacts on $r_-$ rather quickly from $R(v) > r_-$ and then essentially keeps its position close to the Cauchy horizon.

\begin{figure}[t!]
	\includegraphics[width = 0.48\textwidth]{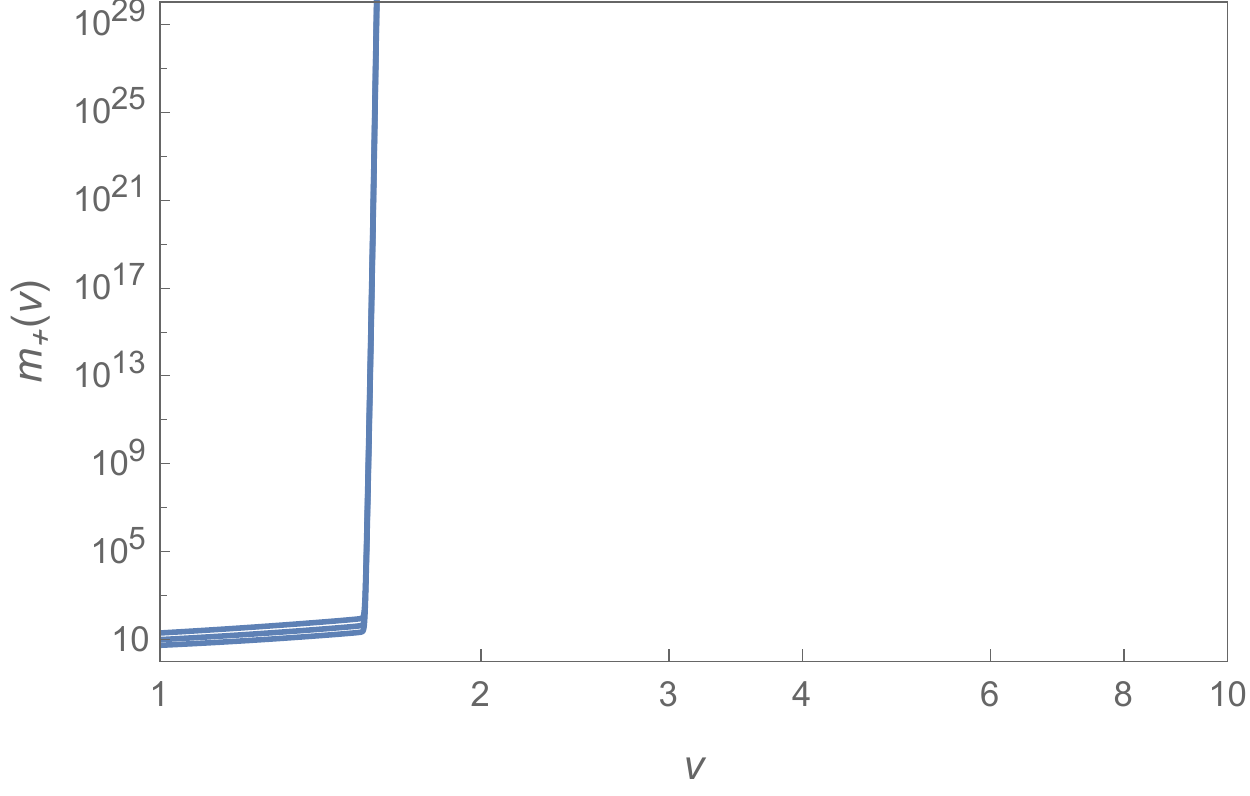} \, 
	\includegraphics[width = 0.48\textwidth]{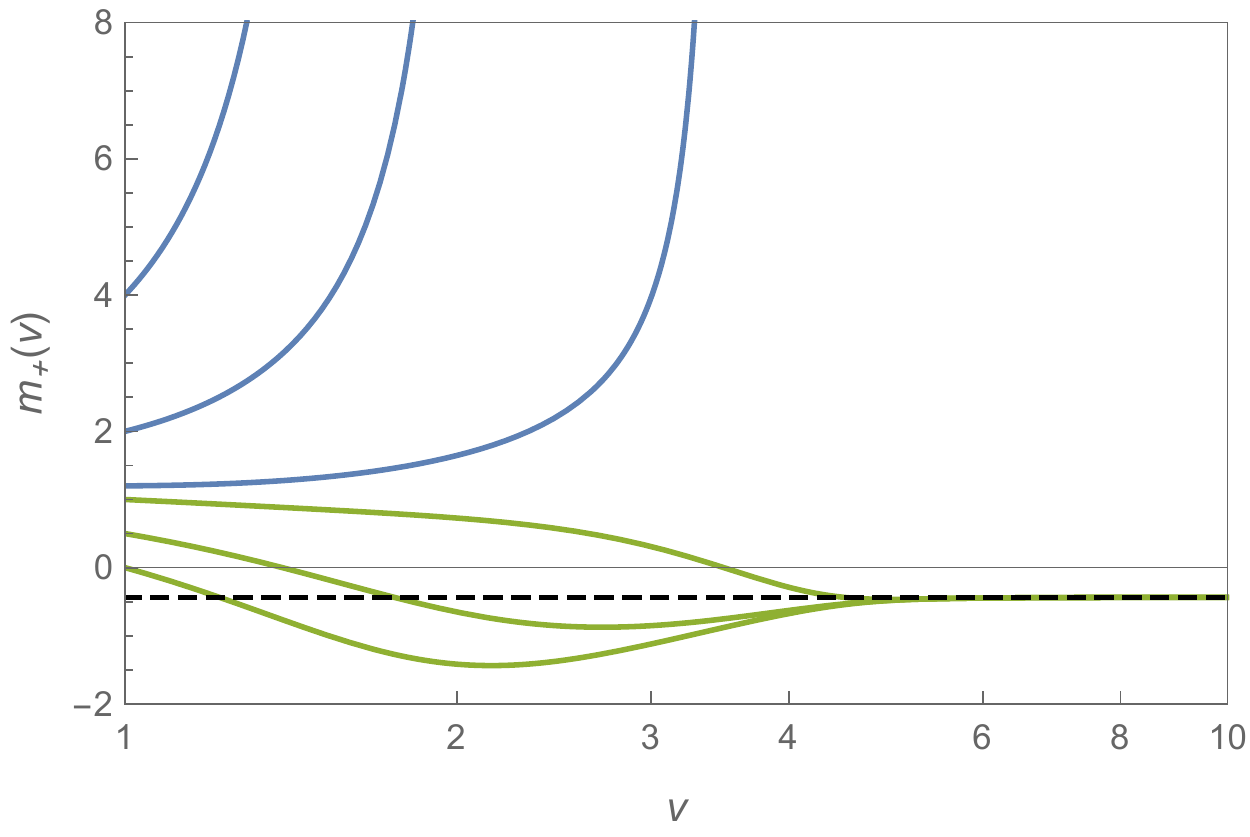} \\
	\includegraphics[width = 0.48\textwidth]{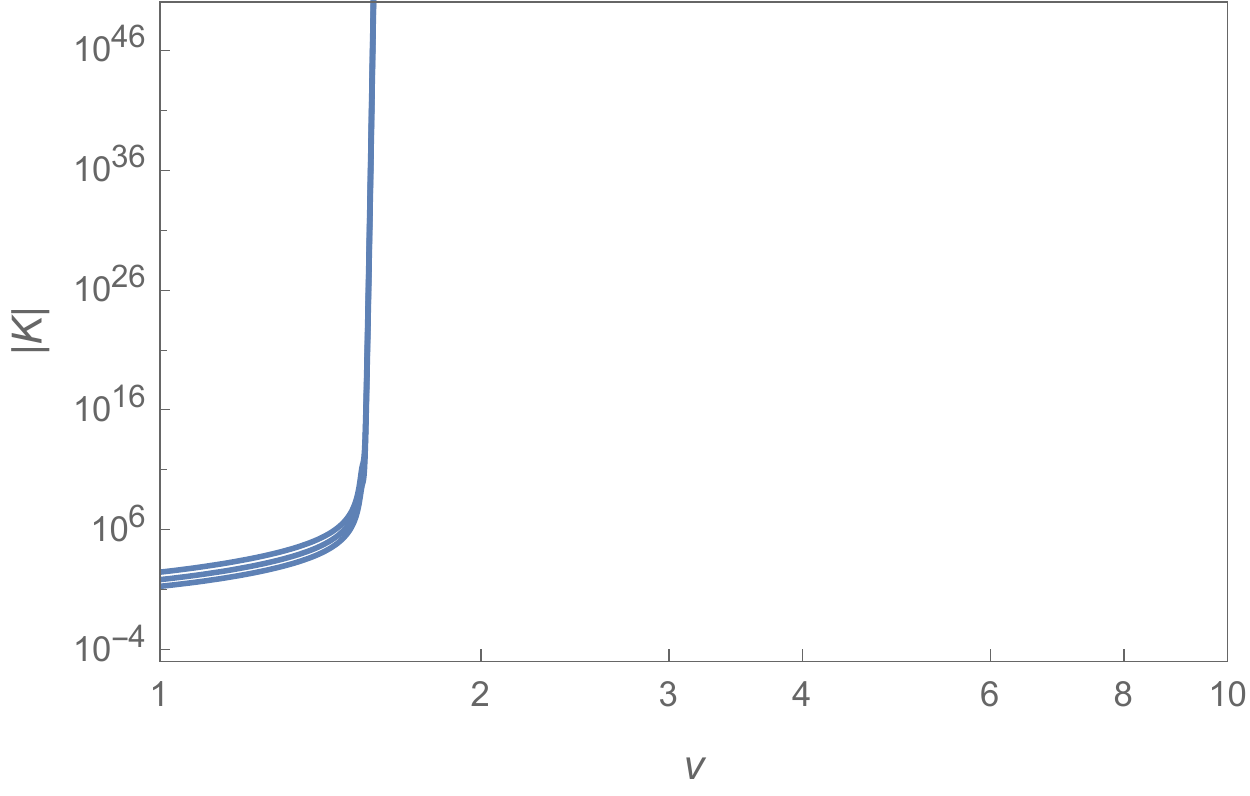} \, 
	\includegraphics[width = 0.48\textwidth]{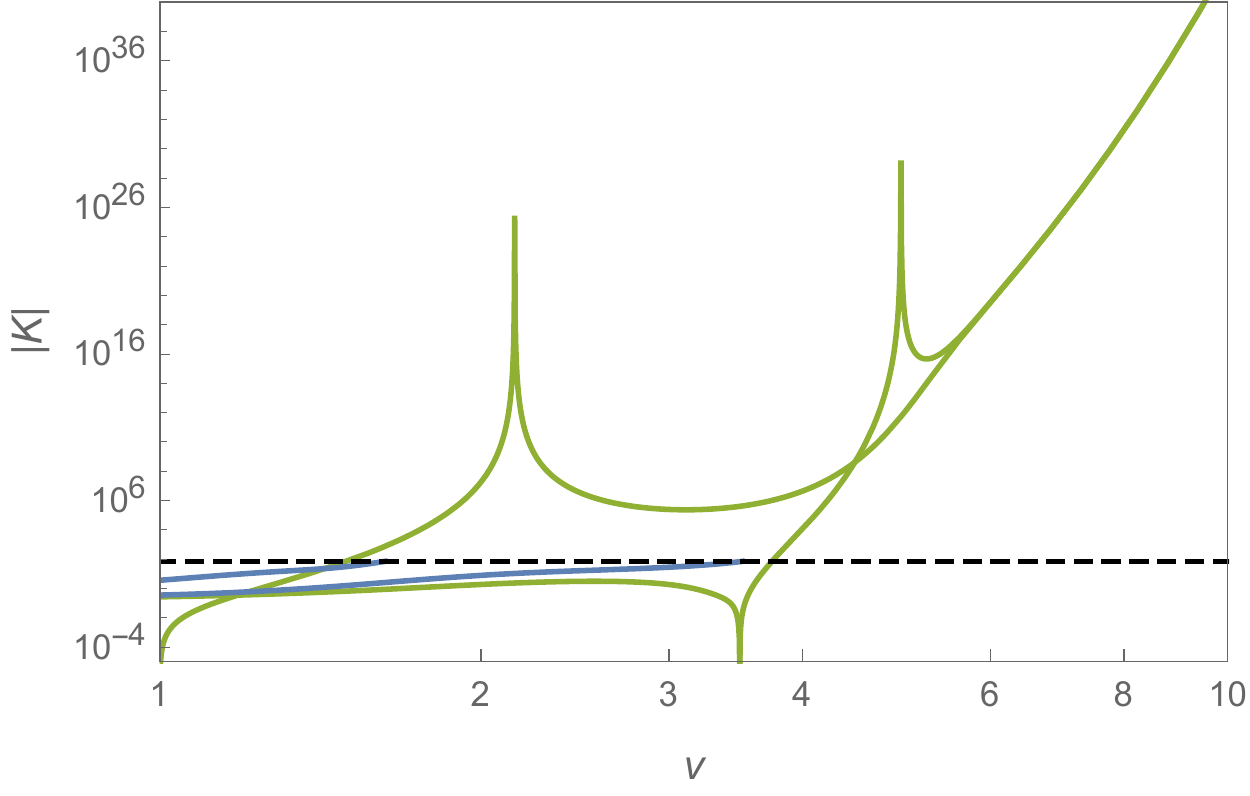} \\
	\caption{Top: Illustration of the mass function $m_+(v)$ inside the shell for a static Reissner-Nordstr{\"o}m (left) and Hayward black hole (right). For the blue lines the numerical integration terminates at finite value $v$ while the green lines reach the late-time attractor of the static Hayward geometry where $m_+^\infty \equiv \lim_{v \rightarrow \infty} m_+(v)$ remains finite (black dashed line). Botton: Illustration of the Kretschmann scalar evaluated at the position of the shell. For the green solutions, the curvature scalar grows polynomially in $v$. The dashed line added in the Hayward case gives the value of $|K|$ at the point where the blue solutions terminate, indicating that the curvature scalar remains finite. The model parameters $Q$ and $l$ have been chosen such that the critical mass of the remnant is $m_{cr} = 1$.}
	\label{fig:s2}      
\end{figure}
Based on the asymptotic solutions for $m_-(v)$ and $R(v)$ one then readily deduces the asymptotic behavior of $F(v)$,
\be\label{eq.s9}
F(v) \simeq - \frac{1}{2} \, r_- \, \kappa_- \, , 
\ee
with $\kappa_- > 0$ the surface gravity at the Cauchy horizon. Remarkably, the negative constant appearing in this relation is universal in the sense that it is the same for both geometries. Eq.\ \eqref{eq.s9} then allows to conclude the asymptotic behavior of $m_+(v)$. It is at this point, where the different structures of $p(m_+)$ enter, yielding
\be\label{eq.s10}
\begin{split}
	{\rm RN:} \qquad & m_+(v) \simeq c_1 \, e^{\kappa_- v} \, v^{-(p-1)} \, , \\
	{\rm H:} \qquad & m_+(v) \simeq -\frac{r_-^3}{2 l^2} \, . 
\end{split}
\ee
In the case of $p(m_+)$ being linear, $m_+(v)$ grows exponentially in $v$, justifying the terminology ``mass-inflation''. For quadratic polynomials $p(m_+)$ admits a second class of solutions where $m_+(v)$ takes a finite, negative value asymptotically. This value is given by the negative root of $p(m_+)$ and thus constitutes a fixed point of \eqref{eq.s4}. The role of this attractor is illustrated at the level of numerical solutions in the top row of Fig.\ \ref{fig:s2}. Imposing initial conditions at $v=1$ (which may be outside of the validity of the Ori-model) one finds that solutions for the Reissner-Nordstr{\"o}m geometry terminate at finite value $v$ (blue curves). In addition to this behavior, the Hayward geometry admits a second class of solutions (green curves): these extend to asymptotically late times and approach the attractor \eqref{eq.s10}.

In order to understand the physics consequences entailed by the late-time behavior of the mass-function, it is instructive to study the growth of the Kretschmann scalar \eqref{eq.curvature} evaluated at the position of the shell. For the asymptotic solutions \eqref{eq.s10} this yields
\be\label{eq.s11}
\begin{split}
	{\rm RN:} \qquad & K|_\Sigma \propto e^{2 \kappa_- v} \, v^{-2 (p-1)} \, , \\
	{\rm H:} \qquad & K|_\Sigma \propto v^{6(p-1)} \, . 
\end{split}
\ee
Thus the structure of $p(m_+)$ \emph{can lead to a significant weakening of the mass-inflation effect for regular black holes}. The origin of the growths \eqref{eq.s11} in the two cases is quite different though: while the exponential growths in the Reissner-Nordstr{\"o}m case is directly related to the growths of the mass-function, the polynomial growths in the Hayward case is tracked back to the fact that the quantity $(R^3 + 2 l^2 m_+)$, which vanishes at the late-time attractor, also appears in the denominator of the Kretschmann scalar. The scaling laws \eqref{eq.s11} are illustrated in the bottom row of Fig.\ \ref{fig:s2}. For the blue solutions $K|_\Sigma$ diverges at finite values $v$ without reaching the late-time attractor. This is different for the green solutions where the numerical integration confirms the polynomial growths of $K|_\Sigma$ in the Hayward case.

The weakening of the curvature singularity \eqref{eq.s11} may have profound consequences for the geodesic completeness of spacetime. Solving the geodesic equation for a massive, radially free-falling observer in the static, non-critical black hole spacetime  shows that $v=\infty$ can be reached in a finite amount of the observer's proper time. Technically, the new late-time attractor turns the mass-inflation singularity into a weak singularity with respect to both the Tipler and the Kr\'{o}lak definition. This may open the possibility to extend geodesics beyond the singularity. While this is certainly an exciting possibility, this point is currently still awaiting its final clarification.

\subsection{The Ori-model including Hawking radiation}
\label{sect.42}
When discussing the thermodynamics of regular black holes in Sect.\ \ref{sect.32}, we argued that the final state of the evaporation process is a cold remnant with finite mass $m_{cr}$. The mass-inflation effect associated with the Cauchy horizon then raises the question whether this picture is robust against perturbations. The analysis for static, regular black holes showed that there are mechanisms which tame the growth of the curvature singularity, delaying the growths of the Misner-Sharp mass to asymptotically late times. This suggests that the Hawking effect can influence the dynamics. The analysis based on the Ori-model \cite{Bonanno:2022jjp}, summarized in this section, indeed confirms this expectation.

Upon including the Hawking effect, the mass function of the black hole is no longer constant but turns into a $v$-dependent function. The time-dependence can be determined from the mass-loss formula \eqref{massloss}. Following the strategy of the previous section, focusing on the late-time behavior of the solutions, it is straightforward to determine the leading terms from the Frobenius method
\be\label{eq.h1}
\begin{split}
	{\rm RN:} \qquad & \, m_{\rm Hawking}(v) \simeq m_{cr} \left(1 + 2 \, \frac{15 \pi \, m_{cr}^3}{ v} + 48 \left( \frac{15 \pi m_{cr}^3}{v} \right)^{3/2} + \cdots \right) \, , \\
	{\rm H:} \qquad & \, m_{\rm Hawking}(v) \simeq m_{cr} \left( 1 + 6 \, \frac{80 \pi \, m_{cr}^3}{v} + 80 \, \left( \frac{80 \pi \, m_{cr}^3}{v} \right)^{3/2}  + \cdots \right) \,  . 
\end{split}
\ee
Here the dependence of $m(v)$ on $Q$ and $l$ are encoded in $m_{cr}$. The expansion shows that $\lim_{v \rightarrow \infty} m(v) = m_{cr}$, giving the expected mass for the cold remnant.  Remarkably, the late-time behavior is universal in the sense that there are no free parameters entering the first three terms of the expansion. 

Starting from Price's law \eqref{eq.O10}, we then include the dynamics of the background by $m_0 \mapsto m_{\rm Hawking}(v)$
\be\label{eq.h2}
m_-(v) = m_{\rm Hawking}(v) - \frac{\beta}{(v/v_0)^{p-1}} \, . 
\ee
The expansion \eqref{eq.h1} then entails that the late-time behavior of $m_-(v)$ is actually fixed by the Hawking effect, the contribution of the Price tail being subleading compared to the time-dependence of $m_{\rm Hawking}(v)$. Moreover, the position of the roots of the lapse function $f(r,v)$ is no longer static: one inherits an apparent event horizon $r_+(v)$ and an apparent Cauchy horizon $r_-(v)$ which both approach the critical radius $r_{cr}$ at asymptotically late times.

\begin{figure}[t!]
	\includegraphics[width = 0.48\textwidth]{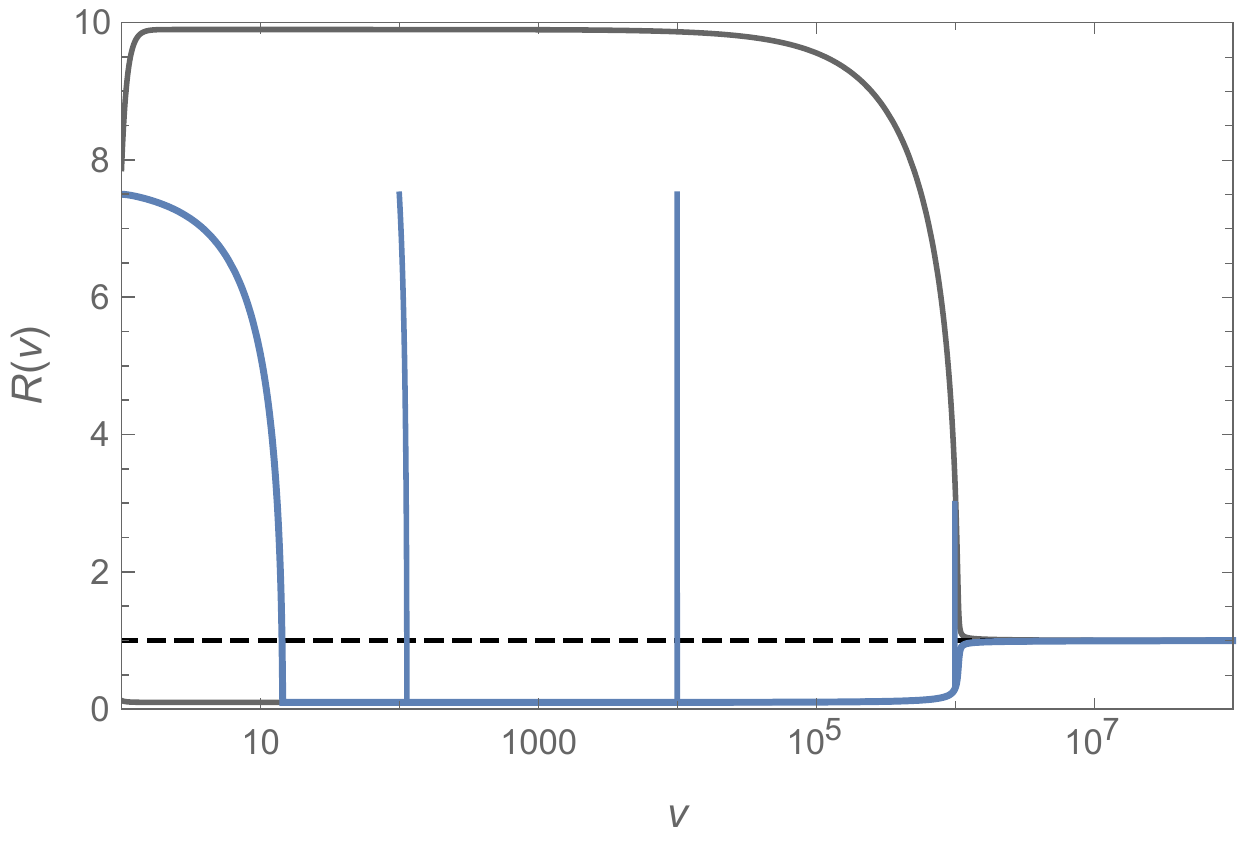} \, 
	\includegraphics[width = 0.48\textwidth]{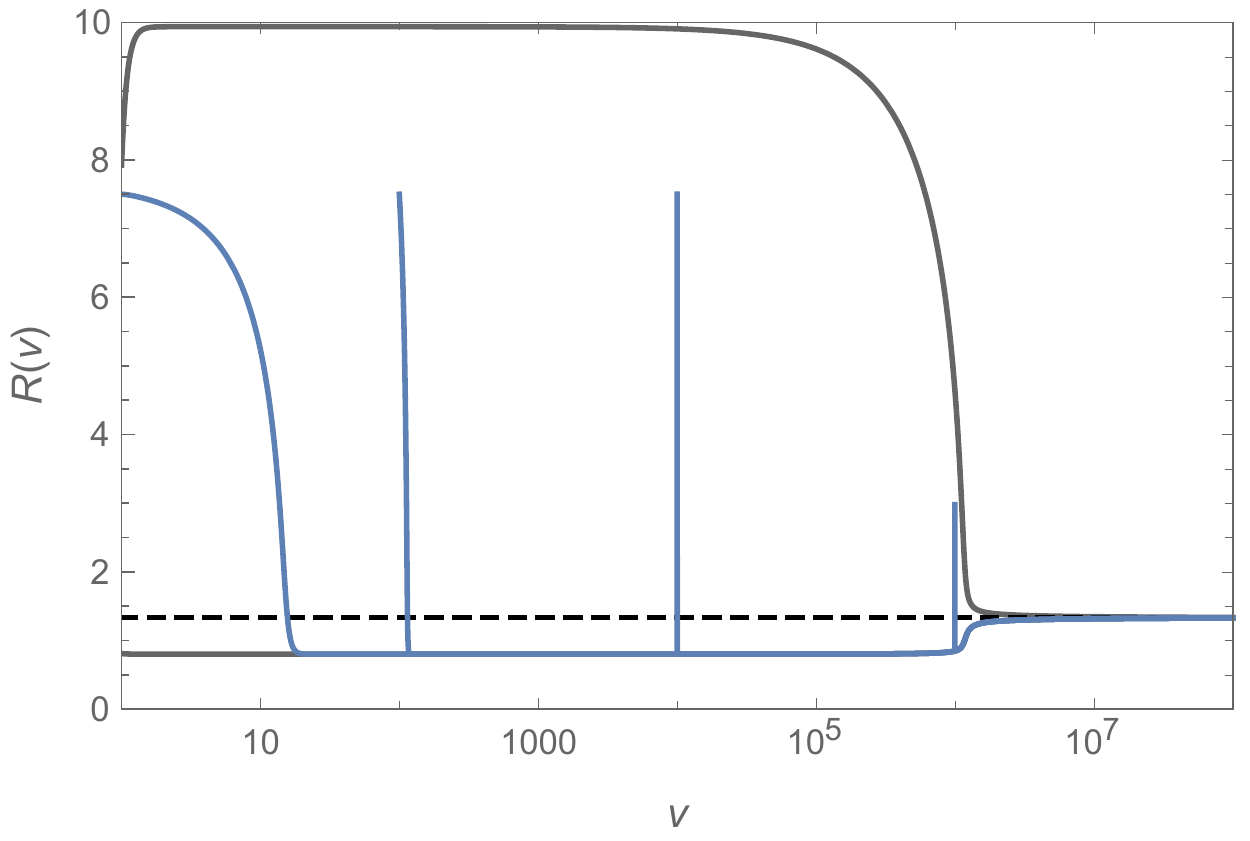}
	\caption{Illustration of the shell dynamics for the background spacetime given by an evaporating Reissner-Nordstr{\"o}m (left) and Hayward black hole (right) (blue lines). The gray lines depict the positions of the apparent event (top) and Cauchy horizons (bottom). In both cases the dashed line depicts $r_{cr}$. The model parameters $Q$ and $l$ have been chosen such that the critical mass of the remnant is $m_{cr} = 1$.}
	\label{fig:6}      
\end{figure}
Having $m_-(v)$ at our disposal, we can again use \eqref{eq.O9} to find the dynamics of the shell. The result obtained from numerical integration is illustrated in Fig.\ \ref{fig:6}. Here the position of the apparent horizons is indicated by the opaque gray lines. In analogy to the static case, a shell starting between the apparent horizons quickly falls towards the apparent Cauchy horizon and subsequently trails the dynamics of $r_-(v)$. This behavior is independent of the initial conditions and details of the geometry.

The crucial difference with the static background then occurs at the level of \eqref{eq.s9}. Including the mass loss one has
\be\label{eq.h3}
\begin{split}
	{\rm RN:} \qquad & \, F(v) \simeq - \left( \frac{15 \pi \, m_{cr}^3}{v} \right)^{1/2} \, , \\
	{\rm H:} \qquad & \, F(v) \simeq - \left( \frac{80 \pi \, m_{cr}^3}{v} \right)^{1/2}  \,  .    
\end{split}
\ee
Thus the function $F(v)$ appearing on the right-hand side of the dynamical equation for $m_+(v)$ vanishes asymptotically. The power-law governing the decay of $F(v)$ is again independent of the geometry under consideration. Heuristically, this can be understood from the fact that the asymptotic geometry is a remnant with vanishing surface gravity. Hence the leading terms in \eqref{eq.s9} vanish and the dynamics of $F(v)$ starts with the subleading order as compared to the static case.

\begin{figure}[t!]
	\includegraphics[width = 0.48\textwidth]{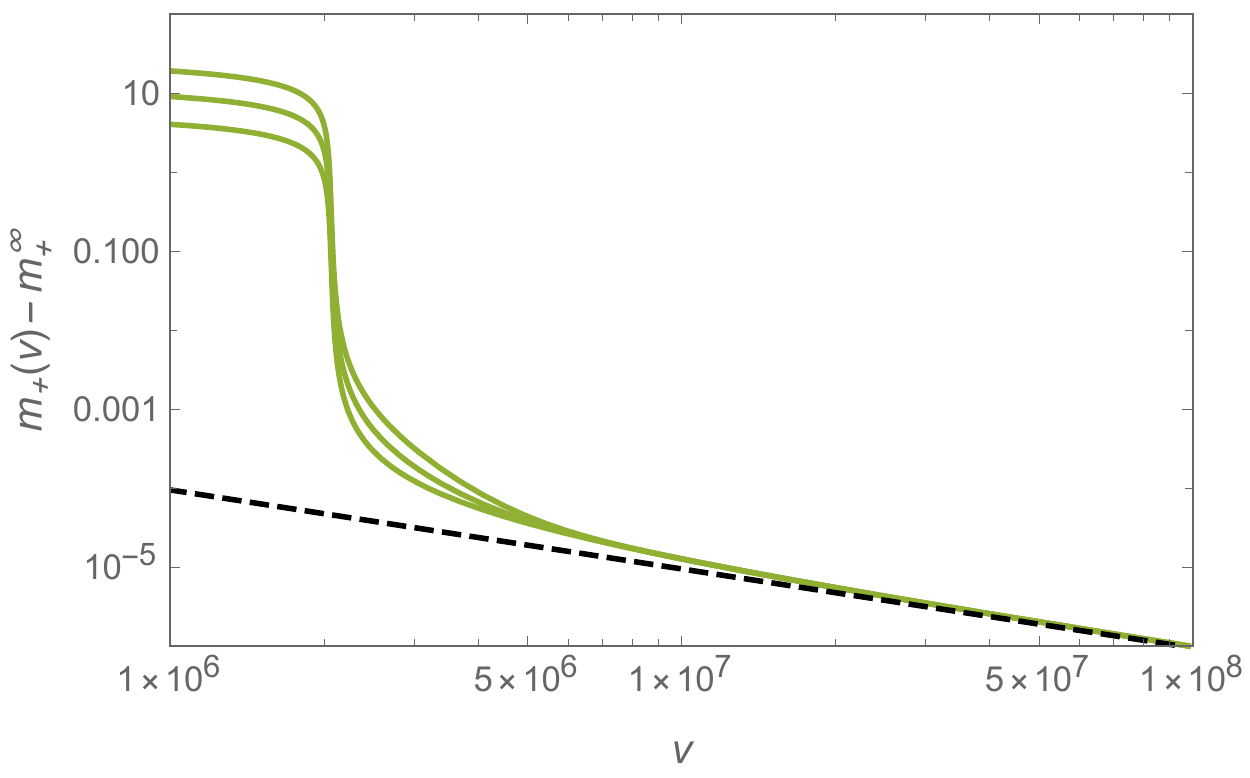} \, 
	\includegraphics[width = 0.48\textwidth]{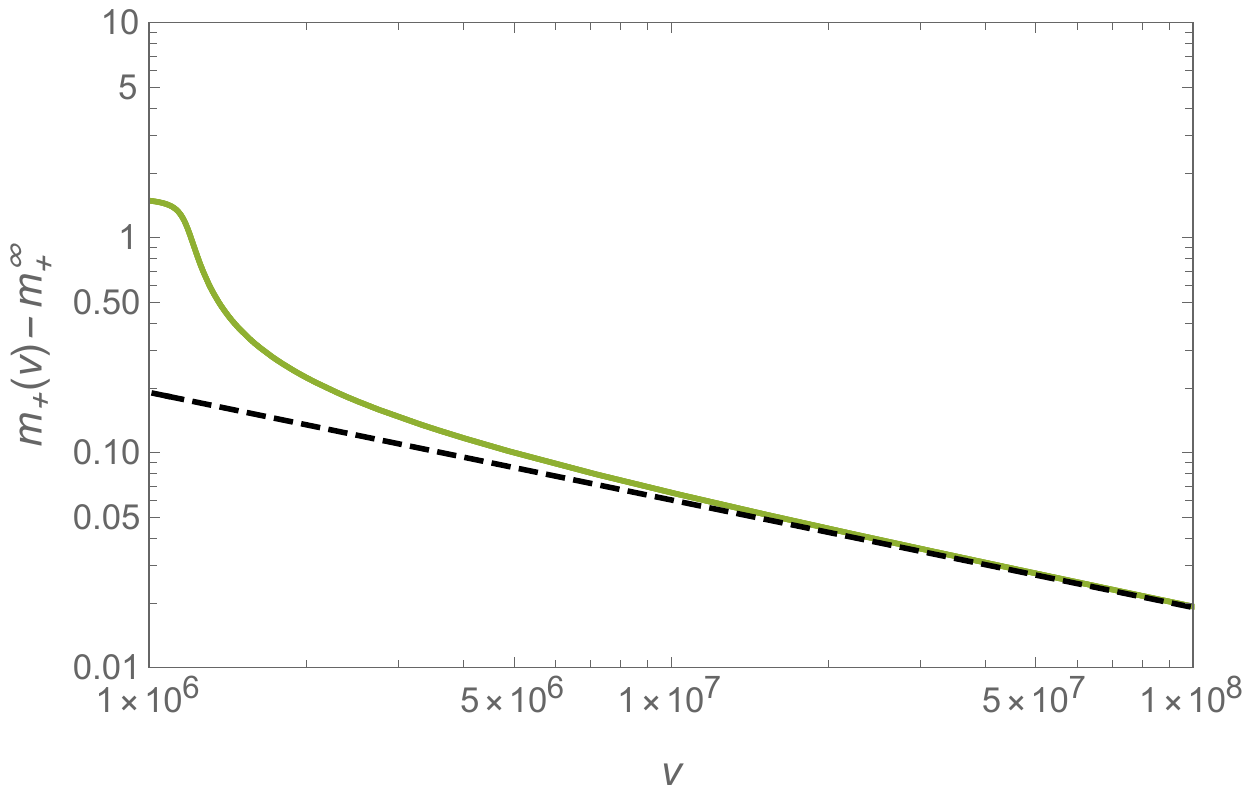} \\
	\includegraphics[width = 0.48\textwidth]{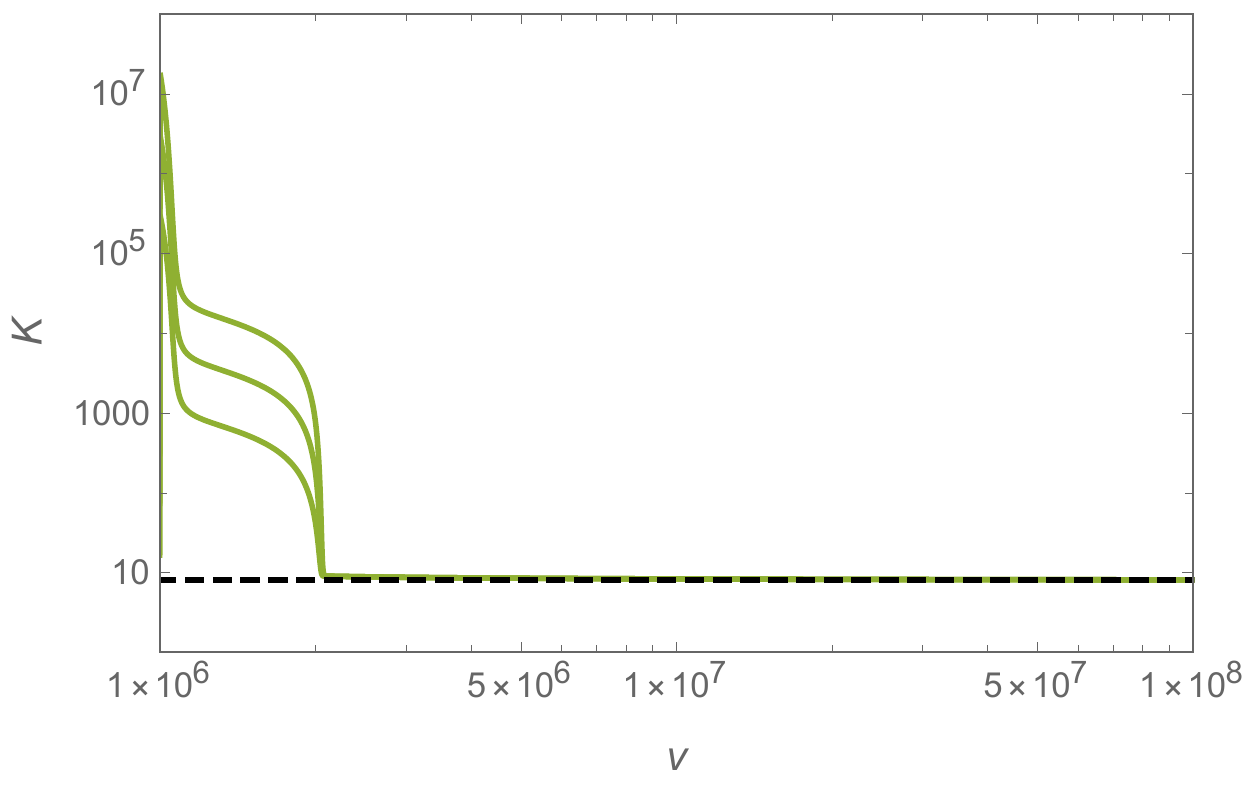} \, 
	\includegraphics[width = 0.48\textwidth]{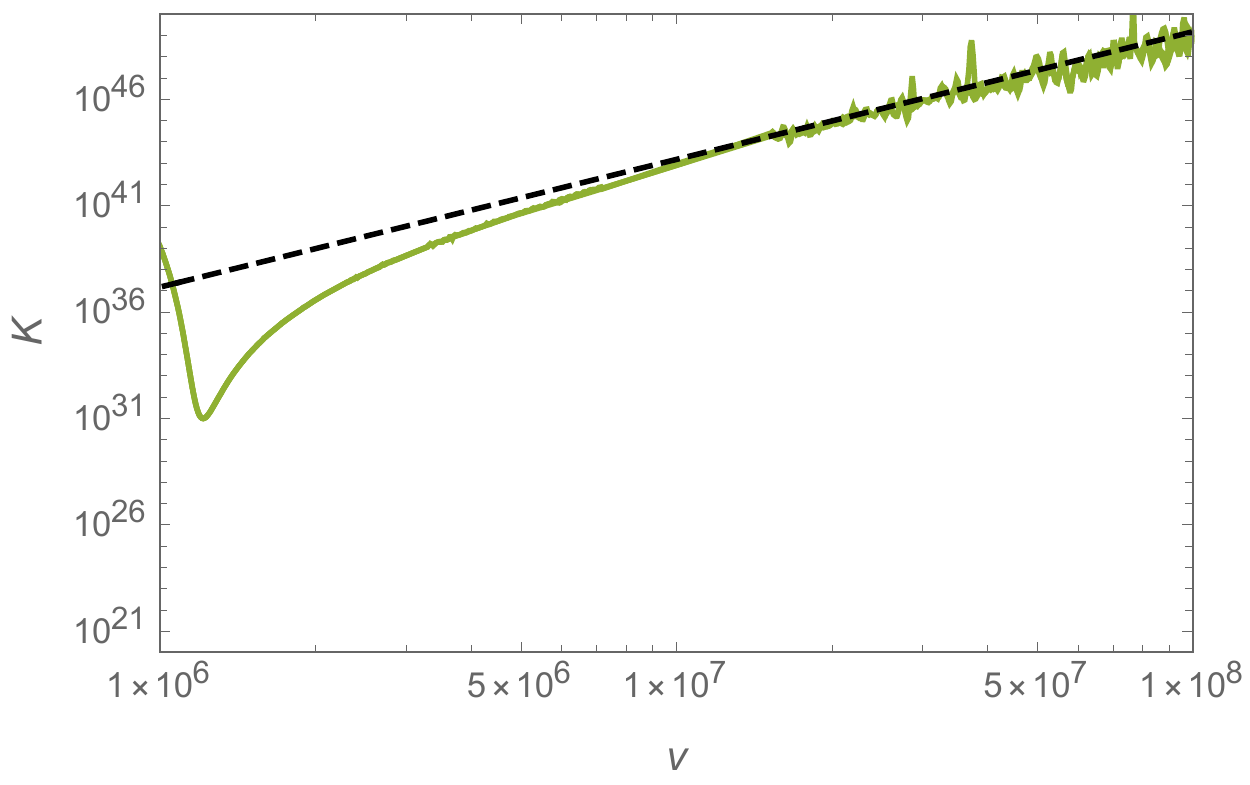} \\
	\caption{Top: Illustration of the mass function $m_+(v)$ inside the shell for a Reissner-Nordstr{\"o}m (left) and Hayward black hole (right) undergoing mass-loss due to Hawking radiation. Initial conditions are imposed at late times. All green lines are attracted to the late-time attractor where $m_+(v)$ approaches the constant $m_+^\infty \equiv \lim_{v \rightarrow \infty}m_+(v)$. Botton: Illustration of the Kretschmann scalar evaluated at the position of the shell. For the Reissner-Nordstr{\"o}m attractor $K|_\Sigma$ remains finite. In the Hayward case, it grows polynomially in $v$. The model parameters $Q$ and $l$ have been chosen such that $m_{cr} = 1$.}
	\label{fig:8}      
\end{figure}
The modification \eqref{eq.h3} has profound consequences for the late-time behavior of $m_+(v)$. In contrast to the static case, the mass function in the interior of the shell approaches a constant value 
\be\label{eq.h4}
\begin{split}
	{\rm RN:} \qquad & m_+(v) \simeq m_{cr} + \frac{30 \pi \, m_{cr}^4}{v} \, , \\
	{\rm H:} \qquad & m_+(v) \simeq - 2 m_{cr} + 12 m_{cr} \, \left( \frac{80 \pi \, m_{cr}^3}{v} \right)^{1/2} \, . 
\end{split}
\ee
Again this new attractor appears in both geometries. The different fall-off behavior in the subleading term for the Reissner-Nordstr{\"o}m case is due to the cancellation of the $v^{-1/2}$-contributions in the series. The approach of numerical solutions to this late-time attractor is illustrated in Fig.\ \ref{fig:8}, top row. The new attractor behavior then also propagates into the Kretschmann scalar evaluated at the position of the shell
\be\label{eq.h4curve}
\begin{split}
	{\rm RN:} \qquad & K|_{\Sigma} \simeq \frac{8}{m_{cr}^4} \, , \\
	{\rm H:} \qquad & K|_{\Sigma} \simeq  \frac{59049 \, v^6}{4096 \, m_{cr}^{10}} \, .  
\end{split}
\ee
The approach of numerical solutions to this attractor is illustrated in the bottom row of Fig.\ \ref{fig:8}. The remarkable feature of this result is that \emph{there is no curvature singularity building up in the Reissner-Nordstr{\"o}m case}. The mechanism underlying the polynomial growths in the Hayward case is identical to the one observed in the static case. A consistent solution for $m_+(v)$ requires the vanishing of the first factor in $p(m_+)$ which then leads to cancellations among leading terms in the denominator of the curvature scalar. As a result, one again experiences a power-law growth of $K|_\Sigma$. 

\begin{figure}[t!]
	\includegraphics[width = 0.48\textwidth]{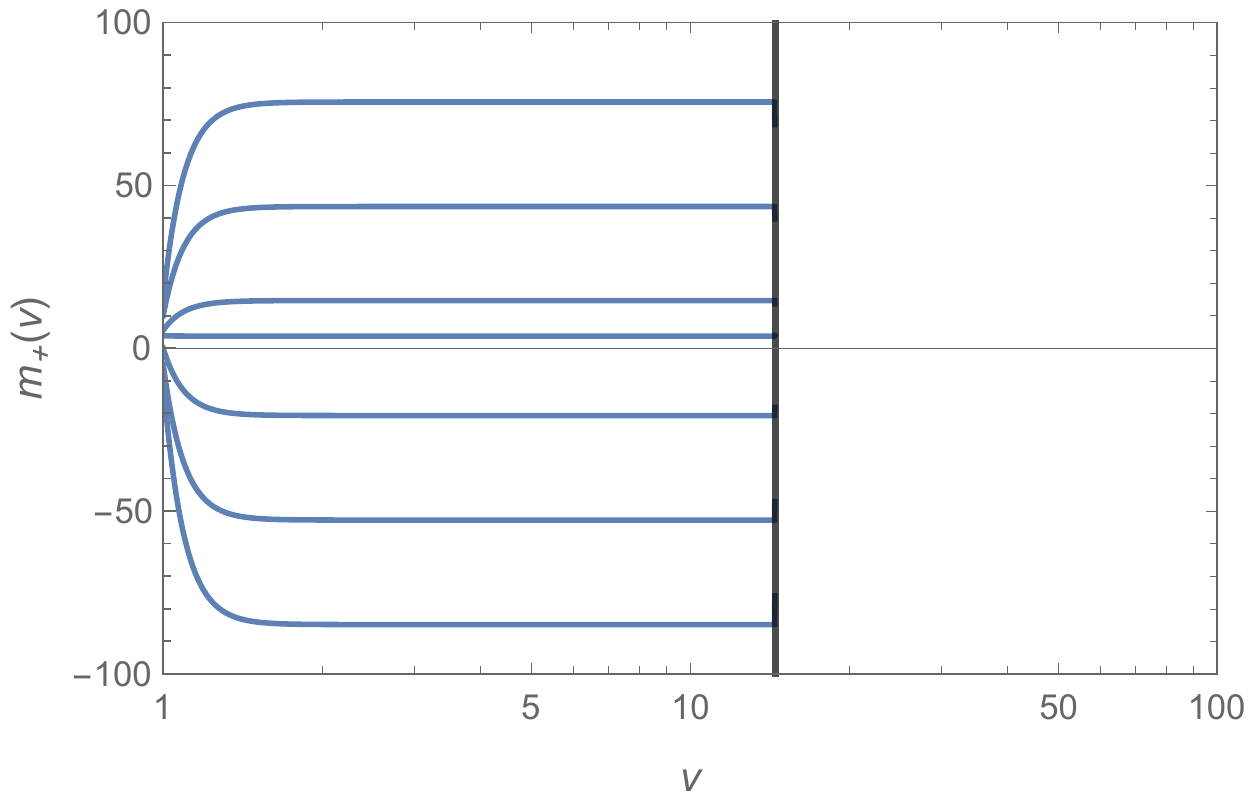} \, 
	\includegraphics[width = 0.48\textwidth]{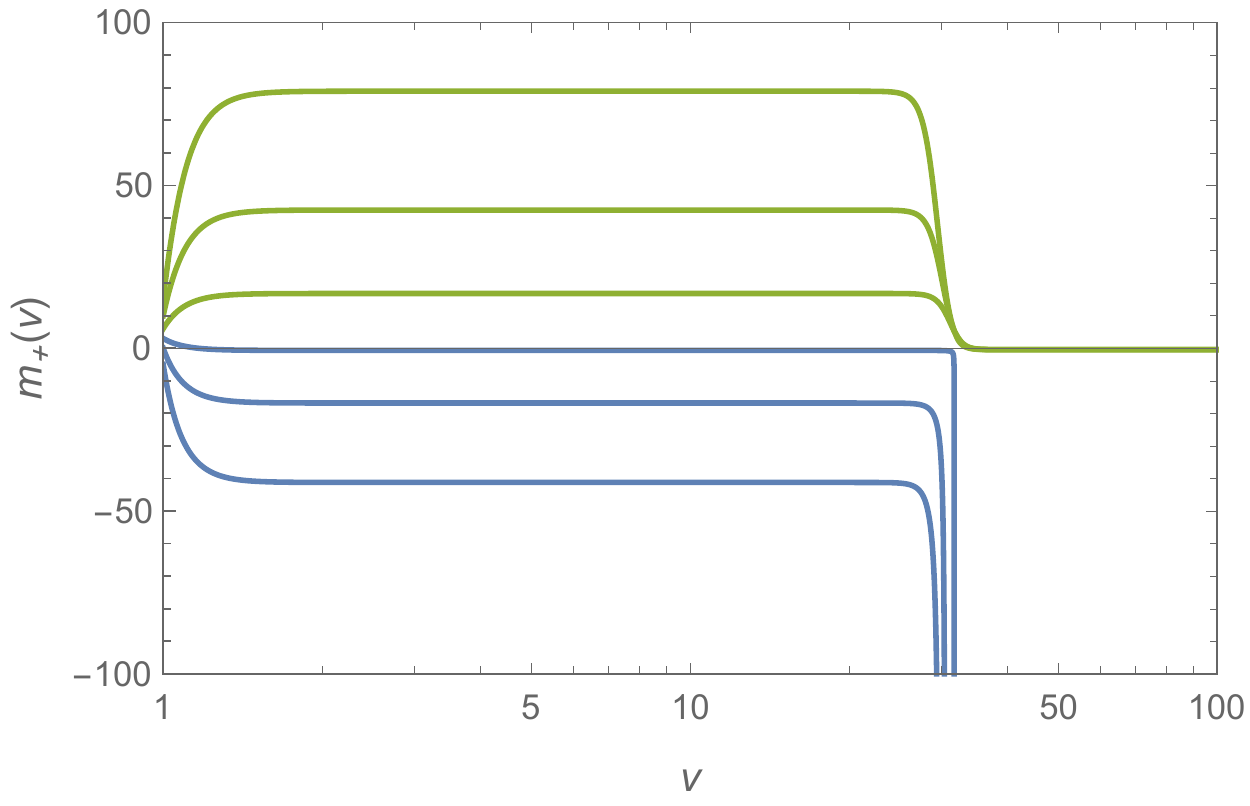}
	\caption{Solutions for the mass function $m_+$ in the inner sector of the shell for the Reissner-Nordstr{\"o}m (left) and Hayward black hole (right). Initial conditions are imposed at ``early times'' $v=1$ and the background is provided by the most left trajectories $R(v)$ shown in Fig.\ \ref{fig:6}. The blue solutions terminate at finite value of $v$. For the Hayward case there exists a critical value for $m_+(v)$. Solutions starting above this critical value are depicted by the green lines and extended up to $v=\infty$. Again $Q$ and $l$ are chosen such that $m_{cr} = 1$.}
	\label{fig:7}       
\end{figure}

An important prerequisite for solutions following the attractor \eqref{eq.h4} is that the initial conditions for the shell are imposed at sufficiently late times, close to the regime where the black hole has almost reached the final phase of its evaporation process, cf.\ Fig.\ \ref{fig:3}. In light of the discussion \cite{Carballo-Rubio:2021bpr,DiFilippo:2022qkl}, it is important to clarify whether this attractor can also be reached from initial conditions imposed at early times, $v=1$ say. While the corresponding analysis is most likely outside of the validity range of the Ori-model (the Price tail and the approximation of the perturbation by a thin shell being reasonable at asymptotically late times only), we give a tentative answer in Fig.\ \ref{fig:7}. This reveals that the static and dynamical background models share the same early-time behavior: for the Reissner-Nordstr{\"o}m black hole all solutions again terminate at finite $v$. For the Hayward case there is again a range of initial conditions where the solutions connect to the asympotic late-time attractors. These are  highlighted by the green lines in the right diagram of Fig.\ \ref{fig:7}.
The existence of these solutions can again be traced back to the quadratic nature of $p(m_+)$ which induces a plateau for $m_+$ even before one enters into the regime where $F(v)$ decays.
%

In order to conclude our discussion, it is important to highlight the relevance of the attractor \eqref{eq.h4} with respect to the geodesic completeness of spacetime. In this context, we stress that the asymptotic geometry is an extremal black hole. Investigating the geodesics of radially free-falling massive observers in such a background one finds that the curvature singularity building up at $v=\infty$ can no longer be reached in a finite amount of the observer's proper time. Hence these observers will not encounter the singularity and questions related to its strength may actually become academic. Whether this result extends to all observers and the full dynamical background is currently still an open question. Nevertheless, the exposition in this section shows that the dynamics of the black hole mass function induced by the emission of Hawking radiation is a crucial element when analyzing the late-time stability of regular black holes. 

\section{Conclusions}
\label{sect.5}
Static black hole solutions in general relativity are characterized by curvature singularities hidden behind an event horizon. The singularities are often taken as a signal for the breakdown of the classical theory which should be removed by quantum (gravity) effects. A common strategy anticipating such an effect replaces the singular part of the black hole spacetime by a regular patch of de Sitter space. In this way one arrives at regular black holes satisfying the limited curvature hypothesis \cite{ansoldi}. Examples include Bardeen-type black holes \cite{bardeen1968non,Ayon-Beato:2000mjt}, the Hayward geometry \cite{hayward}, RG-improved black holes \cite{Bonanno:2000ep} and Planck stars \cite{Rovelli:2014cta}.

A direct consequence of the de Sitter core is that asymptotically flat regular black holes must have (at least) two horizons, an outer event horizon and an inner Cauchy horizon. Thus, in their simplest incarnation they have the same horizon structure as a charged Reissner-Nordstr{\"o}m black hole. The appearance of the Cauchy horizon entails drastic consequences for the black hole evaporation process due to the emission of Hawking radiation: instead of evaporating completely within a finite time-span, the final state of a spherically symmetric regular black hole is a cold, regular remnant corresponding to an extremal black hole.

The similarity to the Reissner-Nordstr{\"o}m geometry suggests that regular black holes may suffer from a dynamical instability, the so-called mass-inflation effect. In brief the effect states that a tiny perturbation crossing the event horizon and impacting on the Cauchy horizon induces a curvature singularity in its interior. Extrapolating this effect from a static, non-extremal Reissner-Nordstr{\"o}m black hole to regular black holes then suggests that the instability can reintroduce spacetime singularities dynamically.

The analysis of the mass-inflation effect based on the Ori-model reveals that this analogy comes with significant limitations though \cite{Bonanno:2020fgp,Bonanno:2022jjp}. At the level of static, non-extremal black holes the equation controlling the dynamics of the mass-inflation effect \emph{is structurally different} for the Reissner-Nordstr{\"o}m geometry and regular black holes of the Hayward and RG-improved type. As a consequence, the latter admit solutions where the mass-function remains constant and the curvature in the inner sector of the shell grows polynomial in the ingoing Eddington-Finkelstein coordinate $v$ 
only.\footnote{This conclusion has been challenged in \cite{Carballo-Rubio:2021bpr,DiFilippo:2022qkl} (also see \cite{Carballo-Rubio:2018pmi} for an earlier analysis based on the DTR-relations), claiming that the onset of the attractor behavior is preceded by a ``fatal'' phase of exponential growth. This conclusion is flawed for three reasons though. Firstly, it builds on an analysis of the system at ``early times'' outside the region of validity of the underlying assumptions: Price's law holds at asymptotically late times only where the optical 
	geometric limit is valid and one is allowed to neglect the ``finite size" 
	effects of the inner potential barrier.
	In a realistic collapse the early-time dynamics is significantly more complicated because of the
	presence of the flux from the collapsing star.  Secondly, closing  eyes to this difficulty and extrapolating the model to early times, Fig.\ \ref{fig:s2} establishes that the analysis of \cite{Carballo-Rubio:2021bpr,DiFilippo:2022qkl} is incomplete. There are initial conditions which actually reach the salient late-time attractor also from early times. 
	Determining the precise initial data requires the analysis of the full dynamical process and is beyond the scope of  a simplified model whose fundamental limitation is the assumption that the Cauchy Horizon always exist at $v=\infty$. Thirdly, a divergence in the asymptotic mass $m_+$ may just not be fatal for the geometry as the Misner-Sharp mass controlling the curvature of spacetime may remain finite. This is clear from eq.\ \eqref{eq.s1} which shows that the limit $m\rightarrow \infty$ has different physical implications for the Reissner-Nordstr{\"o}m and Hayward geometry. Most remarkably, very similar conclusions have already been reached when analyzing the mass-inflation effect for loop black holes \cite{Brown:2011tv}.}

The static analysis is modified significantly once the mass-loss due to Hawking radiation is taken into account \cite{Bonanno:2022jjp}. This leads to two novel types late-time attractors where the curvature either grows polynomially in $v$ (Hayward, RG-improved) or remains finite (Reissner-Nordstr{\"o}m, Bardeen). Moreover, the fact that the final state is an extremal black hole suggests that no observer can actually reach these singularities in a finite proper time. These features are important theoretical prerequisits when trying to establish regular black holes as valid alternatives to the black holes from general relativity.

The existence of late-time attractors taming or even expelling the mass-inflation effect for regular black holes is highly encouraging. This raises the crucial question whether the full dynamics actually reaches these salient regimes. Settling this question may require a full-fledged numerical analysis beyond the analytic models describing the mass-inflation effect at late times. Making this connection will again be an important step towards establishing regular black holes as valid alternatives to the black holes described by general relativity.

\begin{acknowledgement}
	We thank our collaborators N.\ Alkhofer, J.\ Daas, M.\ Galis, G.\ d'Odorico, I.\ van der Pas, A.\ Platania, A.\ Khosravi, B.\ Koch,, S.\ Silveravalle, F.\ Vidotto, M.\ Wondrak, and, foremost,  M.\ Reuter
	for many inspiring discussions developing our understanding of spacetime singularities, regular black holes, and the stability properties of Cauchy horizons. The work of F.S.\ is supported by the Dutch Black Hole Consortium.
\end{acknowledgement}

	\bibliographystyle{jhep}
	\bibliography{generalbib}
	
\end{document}